\documentclass[draft,onecolumn,12pt]{IEEEtran}
\usepackage{enumerate,cite,amsmath,amsfonts,amssymb,subfigure,epsfig,times,graphicx,psfrag}
\usepackage{float}
\usepackage{verbatim,color}
\psfull
\newcommand{\mbf}{\boldmath}
\newcommand{\smbf}{\small \boldmath}

\def\sbh{{\mbox {\smbf $h$}}}

\def\b0{{\mbox {\mbf $0$}}}
\def\bA{{\mbox {\mbf $A$}}}

\def\bh{{\mbox {\mbf $h$}}}

\def\bs{{\mbox {\mbf $s$}}}

\def\bw{{\mbox {\mbf $w$}}}
\def\bx{{\mbox {\mbf $x$}}}
\def\bv{{\mbox {\mbf $v$}}}

\def\bS{{\mbox {\mbf $S$}}}
\def\sbv{{\mbox {\smbf $v$}}}
\def\sbw{{\mbox {\smbf $w$}}}
\def\sbS{{\mbox {\smbf $S$}}}

\def\bU{{\mbox {\mbf $U$}}}
\def\bI{{\mbox {\mbf $I$}}}
\def\bR{{\mbox {\mbf $R$}}}

\def\bQ{{\mbox {\mbf $Q$}}}

\def\bb0{{\mathbf{0}}}

\def\bLambda{\mbox{\boldmath $\Lambda$}}

\def\b_beta{\mbox{\boldmath $\beta$}}
\def\hb_beta{\mbox{\boldmath $\hat \beta$}}

\def\bPhi{\mbox{\boldmath $\Phi$}}

\def\th{{t}}

\newtheorem{theorem}{Proposition}
\newtheorem{Lemma}{Lemma}

\newtheorem{Remark}{Remark}

\begin{document}
\vfill

\title{Robust Cognitive Beamforming With Partial Channel State Information$^{\star}$\\[0.2cm]}
\author{Lan Zhang$^{\dagger}$, Ying-Chang
Liang$^{\ddagger}$, Yan Xin$^{\dagger}$, and H. Vincent Poor$^*$
\thanks{
$^{\star}$The work is supported by the National University of
Singapore (NUS) under Grants R-263-000-314-101, R-263-000-314-112,
and R-263-000-478-112, by a NUS Research Scholarship, and by the
U.S. National Science Foundation under Grants ANI-03-38807 and
CNS-06-25637. This work was done while Y. Xin was visiting Princeton
University.}
\thanks{$^{\dag}$L. Zhang and Y. Xin are with the Department of Electrical and
Computer Engineering, National University of Singapore, Singapore
118622. (email: zhanglan@nus.edu.sg; elexy@nus.edu.sg)}
\thanks{$^{\ddag}$Y.-C. Liang is with Institute of Infocomm Research, A*STAR, 21 Heng Mui Keng
Terrace, Singapore 119613. (email: ycliang@i2r.a-star.edu.sg)}
\thanks{$^{*}$H. V. Poor is with the Department of Electrical Engineering, Princeton University, Princeton, NJ 08544, USA, (email:
poor@princeton.edu).}} \markboth{IEEE TRANSACTIONS ON WIRELESS
COMMUNICATIONS (REVISED)}{}
\renewcommand{\thepage}{} \maketitle
\begin{center}
{\vspace{2cm}{Suggested Editorial Areas}}:\\[0.5cm]
 Cognitive radio, Multiple-input single-output (MISO), Partial channel state information, Power allocation, Wireless networks.
\end{center}
\renewcommand{\thepage}{} \maketitle
\newpage
\pagenumbering{arabic} \linespread{1.7}
\begin{abstract}
This paper considers a spectrum sharing based cognitive radio (CR)
communication system, which consists of a secondary user (SU) having
multiple transmit antennas and a single receive antenna and a
primary user (PU) having a single receive antenna. The channel state
information (CSI) on the link of the SU is assumed to be perfectly
known at the SU transmitter (SU-Tx). However, due to loose
cooperation between the SU and the PU, only partial CSI of the link
between the SU-Tx and the PU is available at the SU-Tx. With the
partial CSI and a prescribed transmit power constraint, our design
objective is to determine the transmit signal covariance matrix that
maximizes the rate of the SU while keeping the interference power to
the PU below a threshold for all the possible channel realization
within an uncertainty set. This problem, termed the robust cognitive
beamforming problem, can be naturally formulated as a semi-infinite
programming (SIP) problem with infinitely many constraints. This
problem is first transformed into the second order cone programming
(SOCP) problem and then solved via a standard interior point
algorithm. Then, an analytical solution with much reduced complexity
is developed from a geometric perspective. It is shown that both
algorithms obtain the same optimal solution. Simulation examples are
presented to validate the effectiveness of the proposed algorithms.
\end{abstract}
\section{Introduction}\label{section:intro}

One of the fundamental challenges faced by the wireless
communication industry is how to meet rapidly growing demands for
wireless services and applications with limited radio spectrum.
Cognitive radio (CR) technology has been proposed as a promising
solution to tackle such a challenge
\cite{FCC2003,Mitola1999,SimonKaykin:review,Cui07:jsac,Cui07:jsac1,Gastpar_IEEE_TIT_2007,Ghasemi_Sousa_IEEE_TWC_2007,Liang:tradeoff08}.
In a spectrum sharing based CR network, the secondary users (SUs)
are allowed to coexist with the primary user (PU), subject to the
constraint, namely the interference constraint, that the
interference power from the SU to the PU is less than an acceptable
value. Evidently, the purpose of the imposed interference constraint
is to ensure that the quality of service (QoS) of the PU is not
degraded due to the SUs. To be aware of whether the interference
constraint is satisfied, the SUs needs obtain knowledge of the radio
environment cognitively.

In this paper, we consider a spectrum sharing based CR communication
scenario, in which the SU uses a multiple-input single-output (MISO)
channel and the primary user (PU) has one receive antenna. We assume
that the channel state information (CSI) about the SU link is
perfectly known at the SU transmitter (SU-Tx). However, owing to
loose cooperation between the SU and the PU, only the mean and
covariance of the channel between the SU-Tx and the PU is available
at the SU-Tx. With this CSI, our design objective is, for a given
transmit power constraint, to determine the transmit signal
covariance matrix that maximizes the rate of the SU while keeping
the interference power to the PU below a threshold for all the
possible channel realizations within an uncertainty set. We term
this design problem the robust cognitive beamforming design problem.

In non-CR settings, the study of multiple antenna systems with
partial CSI has received considerable attention in the past
\cite{GG02:eigenbeamforming,madhow:partialCSI2001}. Specifically,
the paper \cite{madhow:partialCSI2001} considers the case in which
the receiver has perfect CSI but the transmitter has only partial
CSI (mean feedback or covariance feedback). It was proved in
\cite{madhow:partialCSI2001} that the optimal transmission
directions are the same as those of the eigenvectors of the channel
covariance matrix. However, the optimal power allocation solution
was not given in an analytical form. A universal optimality
condition for beamforming was explored in
\cite{jafar:partialCSI2004}, and quantized feedback was studied in
\cite{Aazhang:quantized03}.

In CR settings, power allocation strategies have been developed for
multiple access channels (MAC) \cite{lan07:jsac} and for
point-to-point multiple-input multiple-output (MIMO) channels
\cite{Liang:jstsp}. Particularly, the solution developed in
\cite{Liang:jstsp} can be viewed as cognitive beamforming since the
SU-Tx forms its main beam direction with awareness of its
interference to the PU. A closed-form method has been present in
\cite{Liang:jstsp}. A water-filling based algorithm is proposed in
\cite{lan07:jsac} to obtain the suboptimal power allocation
strategy. However, the papers \cite{lan07:jsac} and
\cite{Liang:jstsp} assume that perfect CSI about the link from the
SU-Tx to the PU is available at the SU-Tx. Due to loose cooperation
between the SU and the PU, it could be difficult or even infeasible
for the SU-Tx to acquire accurate CSI between the SU-Tx to the PU.

In this paper, we formulate the robust cognitive beamforming design
problem as a semi-infinite programming (SIP) problem, which is
difficult to solve directly. The contribution of this paper can be
summarized as follows.

\begin{enumerate}

\item Several important properties of the optimal solution of the SIP problem, the rank-1 property, and the sufficient and necessary conditions of
the optimal solution, are presented. These properties would
transform the SIP problem into a finite constraint optimization
problem.
\item Based on these properties, we show that the SIP problem can be transformed into
a second order cone programming (SOCP) problem, which can be solved
via a standard interior point algorithm.
\item By exploiting the geometric properties of the optimal solution, a closed-form solution for the SIP problem is also provided.
\end{enumerate}

The rest of this paper is organized as follows. Section
\ref{section:model} describes the SU MISO communication system
model, and the problem formulation of the robust cognitive
beamforming design. Section \ref{section:robust} presents several
important lemmas that are used to develop the algorithms. Two
different algorithms, the SOCP based solution and the analytical
solution, are developed in Section \ref{section:geo} and Section
\ref{section:SDP}, respectively. Section \ref{section:simulation}
presents simulation examples, and finally, Section
\ref{section:conclusion} concludes the paper.

The following notation is used in this paper. Boldface upper and
lower case letters are used to denote matrices and vectors,
respectively, $(\cdot)^H$ and $(\cdot)^{T}$ denote the conjugate
transpose and transpose, respectively, ${\bI}$ denotes an identity
matrix, ${\tt{tr}}(\cdot)$ denotes the trace operation, and
${\tt{Rank}}(\bA)$ denotes the rank of the matrix $\bA$.

\section{Signal Model and Problem Formulation}\label{section:model}
\begin{figure}
      \centering
      \psfrag{hc}{$\bh_s$}
      \psfrag{h}{$\bh\sim\mathcal{CN}(\bh_0,\bR)$}
      \psfrag{PU1}{$\text{PU}$}
      \psfrag{SUtx}{$\text{SU-Tx}$}
      \psfrag{SUrx}{$\text{SU-Rx}$}
      \includegraphics[width = 100mm]{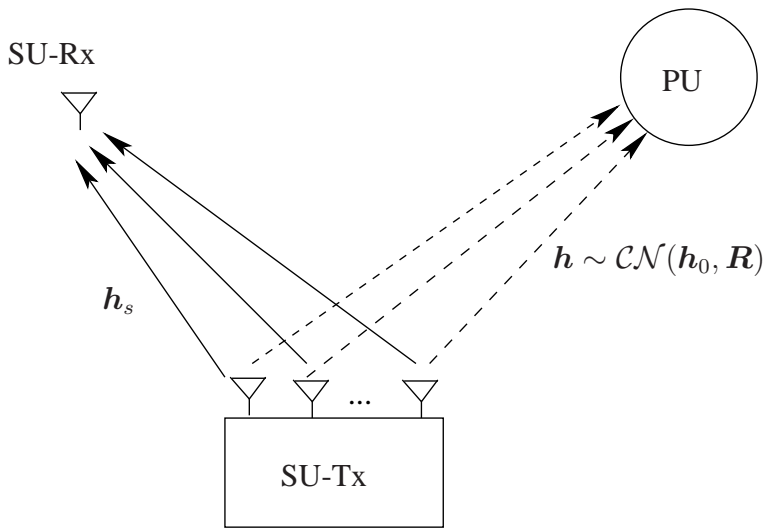}
     \caption{The system model for the MISO SU network coexisting with one PU.}
     \label{fig:sysmodel}
\end{figure}

With reference to Fig. \ref{fig:sysmodel}, we consider a
point-to-point SU MISO communication system, where the SU has $N$
transmit antennas and a single receive antenna. The signal model of
the SU can be represented as
$y=\bh_s^H \bx+n$,
where $y$ and $\bx$ are the received and transmitted signals
respectively, $\bh_s$ denotes the $N\times 1$ channel response from
the SU-Tx to the SU-Rx, and $n$ is independent and identically
distributed (i.i.d.) Gaussian noise with zero mean and unit
variance\footnote{Since the SU receiver cannot differentiate the
interference from the PU from the background noise, the term $n$ can
be viewed as the summation of the interference and the noise. The
variance of $n$ does not influence the algorithms discussed later.
Moreover, the variance of $n$ can be measured at the SU receiver
\cite{lan07:jsac}. }. Suppose that the PU has one receive antenna.
The channel response from the SU-Tx to the PU is denoted by an $N
\times 1$ vector $\bh$. Further, assume that the SU-Tx has perfect
CSI for its own link, i.e., $\bh_s$ is perfectly known at the SU-Tx.
However, due to the loose cooperation between the SU and the PU,
only partial CSI about $\bh$ is assumed to be available at the
SU-Tx. We assume that $\bh_0$ and $\bR$ are the mean and covariance
of $\bh$, respectively\footnote{Due to the cognitive property, we
assume that the SU can obtain the pilot signal from the PU, and thus
can detect the channel information from the PU to the SU. Moreover,
since the SU shares the same spectrum with the PU, based on the
channel from the PU to the SU, the statistics of the channel from
the SU to the PU can be obtained \cite{Liang:covariance01}.
Therefore, we can assume that $\bh_0$ and $\bR$ are known to the
SU.}. In previous work
\cite{madhow:partialCSI2001,Jafar:uniview07,Liang:covariance01,Boche:pcsi03},
partial CSI has been considered in two extreme cases in a non-CR
setting. One is the mean feedback case, $\bR=\sigma^2\bI$, where
$\sigma^2$ can be viewed as the variance of the estimation error;
and the other is the covariance feedback case, where $\bh_0$ is a
zero vector. In this paper, we study the case where the SU-Tx knows
both the mean and covariance of $\bh$ in a CR setting.

The objective of this paper is to determine the optimal transmit
signal covariance matrix such that the information rate of the SU
link is maximized while the QoS of the PU is guaranteed under a
robust design scenario, i.e., the instantaneous interference power
for the PU should remain below a given threshold for all the $\bh$
in the uncertain region. Mathematically, the problem is formulated
as follows:
\begin{align}
\begin{split}
\textbf{Robust design problem}~(\mathbf{P1}):&\quad~\max_{\sbS\geq0}~ \log(1+\bh_s^H \bS \bh_s)\\
\text{subject to}:~&{\tt{tr}}(\bS)\leq \bar{P},~\text{and}~ \bh^H
\bS \bh\leq P_{\th}~\text{for}~(\bh-\bh_0)^H\bR^{-1}(\bh-\bh_0)\leq
\epsilon,
\end{split}
\end{align}
where $\bS$ is the transmit signal covariance matrix, $\bar{P}$ is
the transmit power budget, $P_{\th}$ is the interference threshold
of the PU, and $\epsilon$ is a positive constant. The parameter
$\epsilon$ characterizes the uncertainty of $\bh$ at the SU.
According to the definition of the uncertainty in
\cite{bental07:selected}, $\mathbf{P1}$ belongs to a type of
ellipsoid uncertainty problem, i.e., the uncertain parameter $\bh$
is confined in a range of an ellipsoid $\mathcal{H}$, where
$\mathcal{H}:\{\bh|(\bh-\bh_0)^H\bR^{-1}(\bh-\bh_0)\leq \epsilon\}$.
Thus, the optimal solution of problem $\mathbf{P1}$ can guarantee
the interference power constraint of the PU for all the
$\bh\in\mathcal{H}$, and thus the robustness of $\mathbf{P1}$ is in
the {\it worst case} sense \cite{Boyd_optimization_book}, i.e., in
the worst case channel realization, the interference constraint
should also be satisfied. If the primary transmission does not
exist, then the interference constraint is excluded, and thus the
problem reduces to a trivial beamforming problem. Hence, we only
focus on the case where the both PU and SU transmission exist.

\begin{Remark}\label{remark:rotation}
An important observation is that the objective function in problem
$\mathbf{P1}$ remains invariant when $\bh_s$ undergoes an arbitrary
phase rotation.  Without loss of generality, we assume, in the
sequel, that $\bh_s$ and $\bh_0$ have the same phase, i.e.,
$\text{Im}\{\bh_s^H\bh_0\}=0$.
\end{Remark}

Since problem $\mathbf{P1}$ has a finite number of decision variable
$\bS$, and is subjected to an infinite number of constraints with
respect to the compact set $\mathcal{H}$, problem $\mathbf{P1}$ is
an SIP problem \cite{Rembert_opt_book}. One obvious approach for an
SIP problem is to transform it into a finite constraint problem.
However, there is no universal algorithm to determine the equivalent
finite constraints such that the transformed problem has the same
solution as the original SIP problem. In the following section, we
first study several important properties of problem $\mathbf{P1}$,
which would be used to transform the SIP problem into its equivalent
finite constraint counterpart.

\section{Properties of The Optimal Solution}\label{section:robust}

The maximization problem $\mathbf{P1}$ is a convex optimization
problem, and thus has a unique optimal solution. The following lemma
presents a key property of the optimal solution of problem
$\mathbf{P1}$ (see Appendix \ref{section:prof1D} for the proof).

\begin{Lemma}\label{lemma:1D}
The optimal covariance matrix $\bS$ for problem $\mathbf{P1}$ is a
rank-1 matrix.
\end{Lemma}

\begin{Remark}\label{remark:vp}
Lemma \ref{lemma:1D} indicates that beamforming is the optimal
transmission strategy for problem $\mathbf{P1}$, and the optimal
transmit covariance matrix can be expressed as
$\bS_{\text{opt}}=p_{\text{opt}}\bv_{\text{opt}}\bv_{\text{opt}}^H$,
where $p_{\text{opt}}$ is the optimal transmit power and
$\bv_{\text{opt}}$ is the optimal beamforming vector with
$\|\bv_{\text{opt}}\|=1$. Therefore, the ultimate objective of
problem $\mathbf{P1}$ is to determine $p_{\text{opt}}$ and
$\bv_{\text{opt}}$.
\end{Remark}

According to Lemma \ref{lemma:1D}, a necessary and sufficient
condition for the optimal solution of problem $\mathbf{P1}$ is
presented as follows (refer to Appendix \ref{section:profcond} for
the proof).
\begin{Lemma}\label{lemma:cond}
A necessary and sufficient condition for $\bS_{\text{opt}}$ to be
the globally optimal solution of problem $\mathbf{P1}$ is that there
exists an $\bh_{\text{opt}}$ such that
\begin{align}
\bS_{\text{opt}}=\arg\max_{\sbS,p}\log(1+\bh_s^H \bS \bh_s),~
\text{subject to}:~{\tt{tr}}(\bS)\leq p,~0\leq p\leq \bar{P},
~\bh_{\text{opt}}^H \bS \bh_{\text{opt}}\leq
P_{\th},\label{prob:finite}
\end{align}
where
\begin{align}
\bh_{\text{opt}}=\arg\max_{\sbh}\bh^H \bS_{\text{opt}}
\bh,~\text{for}~(\bh-\bh_0)^H\bR^{-1}(\bh-\bh_0)\leq
\epsilon\label{eq:sufficienth}.
\end{align}
\end{Lemma}

\begin{Remark}
The vector $\bh_{\text{opt}}$ is a key element for all
$\bh:~(\bh-\bh_0)^H\bR^{-1}(\bh-\bh_0)\leq \epsilon$, in the sense
that, for the optimal solution, the constraint
$\bh_{\text{opt}}^H\bS\bh_{\text{opt}}\leq P_{\th}$ dominates the
whole interference constraints, i.e., all the other interference
constraints are inactive. Thus, if we can determine
$\bh_{\text{opt}}$, the SIP problem $\mathbf{P1}$ is transformed
into a finite constraint problem \eqref{prob:finite}. It is worth
noting that the problem \eqref{prob:finite} has the same form as the
problem discuss in \cite{Liang:jstsp}, in which the CSI on the link
of the SU and the link between SU-Tx and PU are perfectly known at
the SU-Tx. However, unlike the problem in \cite{Liang:jstsp},
$\bh_{\text{opt}}$ in \eqref{prob:finite} is an unknown parameter.
\end{Remark}

In the following lemma (see Appendix \ref{section:prov2D} for the
proof), the optimal beamforming vector $\bv_{\text{opt}}$ is shown
to lie in a two-dimensional (2-D) space spanned by $\bh_0$ and the
projection of $\bh_s$ into the null space of $\bh_0$. Define
$\hat{\bh}=\bh_0/\|\bh_0\|$ and
$\hat{\bh}_{\perp}=\bh_{\perp}/\|\bh_{\perp}\|$, where
$\bh_{\perp}=\bh_s-(\hat{\bh}^H\bh_s)\hat{\bh}$. Hence, we have
$\bh_s=a_{h_s}\hat{\bh}+b_{h_s}\hat{\bh}_{\perp}$ with
$a_{h_s},b_{h_s}\in\mathbb{R}$.
\begin{Lemma}\label{lemma:twodimension}
The optimal beamforming vector $\bv_{\text{opt}}$ is of the form
$a_v\hat{\bh}+b_v\hat{\bh}_{\perp}$ with $a_v,b_v\in \mathbb{R}$.
\end{Lemma}

\begin{figure}
      \centering
      \psfrag{h0}{$\bh_{0}$}
      \psfrag{h1}{$\bh_{1}$}
      \psfrag{hc}{$\bh_{s}$}
      \psfrag{hopt}{$\bh_{\text{opt}}$}
      \psfrag{pv}{$\sqrt{p}\bv$}
      \psfrag{th}{$\alpha$}
      \psfrag{Pthp}{$\sqrt{P_{\th}/p}$}
      \includegraphics[width = 120mm]{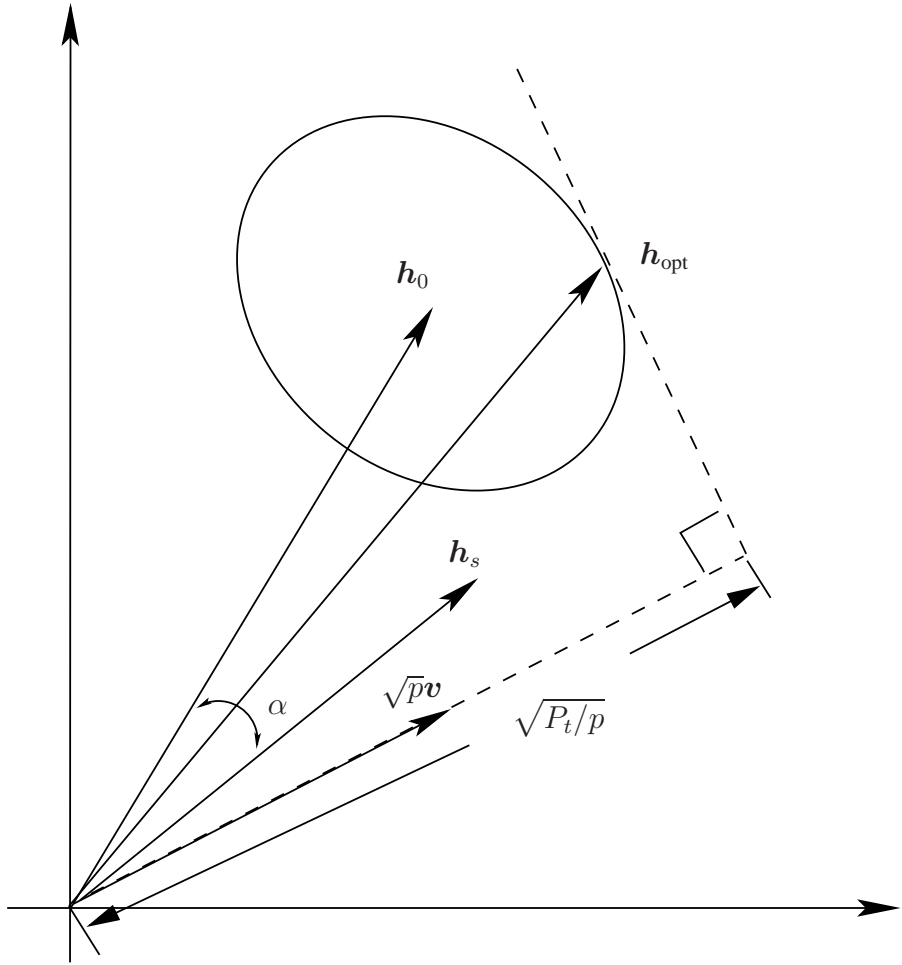}
     \caption{The geometric explanation of Lemma \ref{lemma:twodimension}. The ellipse is the projection of $\bh:=\{(\bh-\bh_0)^H\bR^{-1}(\bh-\bh_0)=\epsilon\}$ on the plane spanned by $\hat{\bh}$ and $\hat{\bh}_{\perp}$.}
     \label{fig:geo3}
\end{figure}

\begin{Remark}
According to Lemma \ref{lemma:twodimension}, we can search for the
optimal beamforming vector $\bv_{\text{opt}}$ on the 2-D space
spanned by $\hat{\bh}$ and $\hat{\bh}_{\perp}$, which simplifies the
search process significantly. The optimal $\bv_{\text{opt}}$ found
in this 2-D space, is also the globally optimal solution of the
original problem $\mathbf{P1}$. As depicted in Fig. \ref{fig:geo3},
problem $\mathbf{P1}$ is transformed into the problem of determining
the beamforming vector $\bv_{\text{opt}}$ in the 2-D space and the
corresponding power $p_{\text{opt}}$. Combining Lemma
\ref{lemma:cond} and Lemma \ref{lemma:twodimension}, it is easy to
conclude that $\bh_{\text{opt}}$ lies in the space spanned by
$\hat{\bh}$ and $\hat{\bh}_{\perp}$.
\end{Remark}

\section{Second Order Cone Programming Solution}\label{section:SDP}

In this section, we solve problem $\mathbf{P1}$ via a standard
interior point algorithm
\cite{Boyd_optimization_book,Luo03:SPbstpaper,Tomluo06:optimization}.
We first transform the SIP problem into a finite constraint problem,
and further transform it into a standard SOCP form, which can be
solved by using a standard software package such as SeDuMi
\cite{Sturm99:sedumi}. One key observation is that if
$\max_{\sbh\in\mathcal{H}(\epsilon)}~\bh^H \bS \bh\leq P_{\th}$,
i.e., the worst case interference constraint of $\mathbf{P1}$ is
satisfied, then the interference constraint of $\mathbf{P1}$ holds.
Combining this observation with Lemma \ref{lemma:1D}, problem
$\mathbf{P1}$ can be transformed as:
\begin{align}
\begin{split}
\textbf{{Equivalent problem ($\mathbf{P2}$):}}\quad
&\underset{p\geq0,\|\sbv\|=1}{\max} \log(1+p\bh_s^H \bv\bv^H \bh_s)\\
\text{subject to}:~&p\leq
\bar{P},~\max_{\sbh\in\mathcal{H}(\epsilon)}~p\bh^H \bv\bv^H \bh\leq
P_{\th},
\end{split}
\end{align}
where $\mathcal{H}(\epsilon):=\{\bh|\bh=\bh_0+\bh_1\}$. It is clear
that maximizing $\log(1+p\bh_s^H \bv\bv^H \bh_s)$ is equivalent to
maximizing $|\sqrt{p}\bh_s^H\bv|$. By defining $\bw=\sqrt{p}\bv$,
the objective function can be rewritten as $|\bh_s^H \bw|$.
Similarly, the interference power can be expressed as $|\bh^H
\bw|^2$. Thus, problem $\mathbf{P2}$ can be further transformed to
\begin{align}\label{eq:wcasesinglevec}
\begin{split}
\max_{\sbw}~&|\bh_s^H \bw|\\
\text{subject to}:~&\|\bw\|\leq
\sqrt{\bar{P}},~\max_{\sbh\in\mathcal{H}(\epsilon)}|\bh^H \bw|\leq
\sqrt{P_{\th}}.
\end{split}
\end{align}

According to the definition of $\mathcal{H}(\epsilon)$, we can
rewrite the worst-case constraint in \eqref{eq:wcasesinglevec} as
\begin{align}
\max_{\sbh\in\mathcal{H}(\epsilon)}|\bh^H \bw|
=\max_{\sbh_1\in\mathcal{H}_1(\epsilon)}|(\bh_0+\bh_1)^H \bw|\le
\sqrt{P_{\th}},
\end{align}
where $\mathcal{H}_1(\epsilon):=\{\bh_1|\bh_1^H\bR^{-1}\bh_1\le
\epsilon\}$. By applying the triangle inequality and the fact that
$\sqrt{\epsilon}\|\bQ\bw\|=\max|\bh_1^H \bw|$ for $\bh_1\in
\mathcal{H}_1(\epsilon)$ (refer to Appendix \ref{appdx:lemma} for
details), the interference power can be transformed as follows:
\begin{align}
|(\bh_0+\bh_1)^H \bw|~ \leq|\bh_0^H \bw|+|\bh_1^H\bw| \leq|\bh_0^H
\bw|+\sqrt{\epsilon}\|\bQ\bw\|,\label{eq:Qineq}
\end{align}
where $\bQ=\mathbf{\Delta}^{-1/2}\bU$ with $\mathbf{\Delta}$ and
$\bU$ being obtained by the eigenvalue decomposition of  $\bR^{-1}$
as $\bR^{-1}=\bU^H\mathbf{\Delta}\bU$. Moreover, since the arbitrary
phase rotation of $\bw$ does not change the value of the objective
function or the constraints, according to Remark
\ref{remark:rotation} and Lemma \ref{lemma:twodimension}, we can
assume that $\bw$, $\bh_s$, and $\bh_0$ have the same phase, i.e.,
\begin{align}\label{eq:image}
\text{Re}\{\bw^H\bh_s\}\geq 0, ~\text{Im}\{\bw^H\bh_0\}=
0,~{\text{and}}~\text{Im}\{\bw^H\bh_s\}= 0.
\end{align}
Hence, the interference constraint can be transformed into two
second order cone inequalities as follows
\begin{align}
\begin{split}\label{eq:SOCconstr}
\sqrt{\epsilon}\|\bQ\bw\|+\bh_0^H \bw\leq \sqrt{P_{\th}},~{\text
{and}}~ \sqrt{\epsilon}\|\bQ\bw\|-\bh_0^H \bw\leq \sqrt{P_{\th}}.
\end{split}
\end{align}
By combining \eqref{eq:wcasesinglevec}, \eqref{eq:SOCconstr}, with
\eqref{eq:image}, problem $\mathbf{P1}$ is transformed into the
standard SOCP problem as follows
\begin{align}\label{eq:SDP}
\begin{split}
&\max_{\sbw} ~\bh_s^H \bw\\
\!\!\!\text{subject to}:\!\|\bw\|\leq
\sqrt{\bar{P}},~\!\text{Im}\{\bw^H\bh_0\}= 0,&~\!
\sqrt{\epsilon}\|\bQ\bw\|+\bh_0^H \bw\leq
\sqrt{P_{\th}},~\!\sqrt{\epsilon}\|\bQ\bw\|-\bh_0^H \bw\leq
\sqrt{P_{\th}}.
\end{split}
\end{align}
Since the parameters $\bh_s$ and $\bh_0$,  and the variable $\bw$ in
\eqref{eq:SDP} have complex values, we first convert them to its
corresponding real-valued form in order to simplify the solution.
Define $\tilde{\bw}:=[\text{Re}\{\bw\}^T,\text{Im}\{\bw\}^T]^T$,
$\tilde{\bh}_0:=[\text{Re}\{\bh_0\}^T,\text{Im}\{\bh_0\}^T]^T$,
$\tilde{\bh}_s:=[\text{Re}\{\bh_s\}^T,\text{Im}\{\bh_s\}^T]^T$,
$\check{\bh}_0:=[\text{Im}\{\bh_0\}^T,-\text{Re}\{\bh_0\}^T]^T$, and
$\tilde{\bQ}:=\left[ \begin{array}{ccc}
\text{Re}\{\bQ\}~-\!\!\!\!\!\!&\text{Im}\{\bQ\}\\
\text{Im}\{\bQ\}~&\text{Re}\{\bQ\}\\
\end{array}\right].$

We then can rewrite the standard SOCP problem \eqref{eq:SDP} as
\begin{align}\label{eq:SDPreal}
\begin{split}
&\max_{\tilde{\sbw}}~ \tilde{\bh}_s^H \tilde{\bw}\\
\!\!\!\text{subject to}:~\!\|\tilde{\bw}\|\leq
\sqrt{\bar{P}},~\!\check{\bh}_0^H\tilde{\bw}=
0,~\!&\sqrt{\epsilon}\|\tilde{\bQ}\tilde{\bw}\|+\tilde{\bh}_0^H
\tilde{\bw}\leq
\sqrt{P_{\th}},~\!\sqrt{\epsilon}\|\tilde{\bQ}\tilde{\bw}\|-\tilde{\bh}_0^H
\tilde{\bw}\leq \sqrt{P_{\th}}.
\end{split}
\end{align}

Problem \eqref{eq:SDPreal} can be solved by a standard interior
point program SeDuMi \cite{Sturm99:sedumi}, which has a polynomial
complexity. In the next section, we develop an analytical algorithm
to solve problem $\mathbf{P1}$, which reduces the complexity of the
interior point based algorithm substantially.

\section{An Analytical Solution}\label{section:geo}

In this section, we present a geometric approach to problem
$\mathbf{P1}$. We begin by studying a special case, the mean
feedback case, i.e., $\bR=\sigma^2\bI$. Due to its special geometric
structure, the mean feedback case problem can be solved via a
closed-form algorithm. We next show that problem $\mathbf{P1}$ can
be transformed into an optimization problem similar to the mean
feedback case. Based on the closed-form solution derived for the
mean feedback case, the analytical solution to problem $\mathbf{P1}$
with a general form of a covariance matrix $\bR$ is presented in
Subsection \ref{subsection:meancov}.

\subsection{Mean Feedback Case}\label{subsection:mean}
Based on the observation in Lemma \ref{lemma:1D} and the definition
of the mean feedback, the special case of problem $\mathbf{P1}$ with
mean feedback can be written as follows.
\begin{align}
\begin{split}
{\textbf{Mean feedback problem ($\mathbf{P3}$): }}\max_{p\ge0,\|\sbv\|=1} &\log(1+p\bh_s^H \bv\bv^H \bh_s)\\
\text{subject to}:~&p\leq \bar{P},~p\bh^H \bv\bv^H \bh\leq
P_{\th},~\text{for}~\|\bh-\bh_0\|^2\leq \epsilon\sigma^2.
\end{split}
\end{align}

Problem $\mathbf{P3}$ has two constraints, i.e., the transmit power
constraint and the interference constraint. Similar to the idea in
\cite{lan07:jsac}, the two-constraint problem is decoupled into two
single-constraint subproblems:
\begin{eqnarray}
\textbf{Subproblem~1~($\mathbf{SP1}$):}&\underset{p\ge0,\|\sbv\|=1}{\max}& ~\log(1+p\bh_s^H \bv\bv^H \bh_s)\\
&\text{subject to}:~&p\leq \bar{P}.\\
\textbf{Subproblem~2~($\mathbf{SP2}$):}&\underset{p\ge0,\|\sbv\|=1}{\max}& ~\log(1+p\bh_s^H \bv\bv^H \bh_s)\\
&\text{subject to}:~&p\bh^H\bv\bv^H\bh\leq
P_{\th},~\text{for}~\|\bh-\bh_0\|^2\leq \epsilon\sigma^2.
\end{eqnarray}

In the sequel, we present the algorithm to obtain the optimal power
$p_{\text{opt}}$ and the optimal beamforming vector
$\bv_{\text{opt}}$ for both subproblems in subsection
\ref{subsection:subsolution}, and describe the relationship between
the subproblems and problem $\mathbf{P3}$ in subsection
\ref{subsection:gloandsub}.

\subsubsection{Solution to subproblems}\label{subsection:subsolution}

For $\mathbf{SP1}$, the optimal power is constrained by the transmit
power constraint, and thus $p_{\text{opt}}=\bar{P}$. Moreover, since
there does not exist any constraints on the beamforming direction,
it is obvious that the optimal beamforming direction is equal to
$\bh_s$, i.e., $\bv_{\text{opt}}=\bh_s/\|\bh_s\|$. Thus, the optimal
covariance matrix $\bS_{\text{opt}}$ for $\mathbf{SP1}$ is
$\bar{P}\bh_s\bh_s^H/\|\bh_s\|^2$. In the following, we focus on the
solution to $\mathbf{SP2}$.

$\mathbf{SP2}$ has infinitely many interference constraints, and
thus is an SIP problem too. By following a similar line of thinking
as in Lemma \ref{lemma:cond}, $\mathbf{SP2}$ can be transformed into
an equivalent problem that has finite constraints (refer to Appendix
\ref{section:profsubcond} for the proof) as follows.
\begin{Lemma}\label{lemma:subcond}
$\mathbf{SP2}$ and the following optimization problem:
\begin{align}
\max_{p\ge0,\|\sbv\|=1}\log(1+p\bh_s^H \bv\bv^H \bh_s),~
\text{subject to}:~p\bh_{\text{opt}}^H \bv\bv^H \bh_{\text{opt}}\leq
P_{\th},\label{prob:finitesub}
\end{align}
where $\bh_{\text{opt}}=\bh_0+\sqrt{\epsilon}\sigma\bv$, have the
same optimal solution.
\end{Lemma}

\begin{figure}
      \centering
      \psfrag{h0}{$\bh_{0}$}
      \psfrag{h1}{$\bh_{1}$}
      \psfrag{hc}{$\bh_{s}$}
      \psfrag{hopt}{$\bh_{\text{opt}}$}
      \psfrag{pv}{$\sqrt{p}\bv_{\text{opt}}$}
      \psfrag{b}{$\beta$}
      \psfrag{th}{$\alpha$}
      \psfrag{Pthp}{$\sqrt{P_{\th}/p}$}
      \includegraphics[width = 100mm]{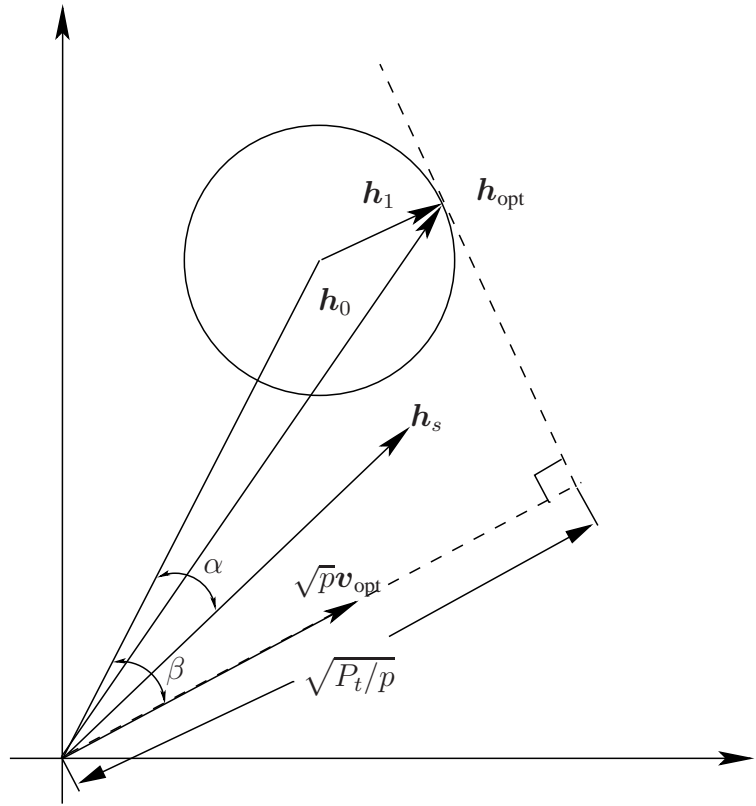}
     \caption{The geometric explanation of problem $\mathbf{P3}$. The circle is the projection of $\bh:=\{\|\bh-\bh_0\|^2=0\}$ on the plane spanned by $\hat{\bh}$ and $\hat{\bh}_{\perp}$.}
     \label{fig:geo}
\end{figure}

According to Lemma \ref{lemma:subcond}, problem
\eqref{prob:finitesub} has the same optimal solution as
$\mathbf{SP2}$. Moreover, according to Lemma
\ref{lemma:twodimension}, the optimal solution $\bv$ of problem
\eqref{prob:finitesub} lies in the plane spanned by $\hat{\bh}$ and
$\hat{\bh}_{\perp}$. We next apply a geometric approach to search
the optimal solution, i.e., by restricting our search space to a 2-D
space. As shown in Fig. \ref{fig:geo}, assume that the angle between
$\bv$ and $\bh_0$ is $\beta$, and the angle between $\bh_s$ and
$\bh_0$ is $\alpha$. It is easy to observe that $0\leq\alpha\leq
\pi/2$\footnote{This follows because if $\alpha\geq \pi/2$, we can
always replace $\bh_s$ by $-\bh_s$ without affecting the final
result, and the angle between $-\bh_s$ and $\bh_0$ is less than
$\pi/2$.}. Since $\bv$ lies in a 2-D space, $\bv$ can be uniquely
identified by the angle $\beta$. Hence, we need only to search for
the optimal angle $\beta_{\text{opt}}$. By exploiting the
relationship between $p$, $\bv$, and $\beta$, the two-variable
optimization problem \eqref{prob:finitesub} can be further
transformed into an optimization problem with a single variable
$\beta$, which can be readily solved.

By observing Fig. \ref{fig:geo}, the angle between $\bh_s$ and $\bv$
is $\beta-\alpha$, and hence the objective function of
\eqref{prob:finitesub} can be expressed as
\begin{align}
\max_{\|\bv\|=1}\log(1+p\bh_s^H\bv\bv^H\bh_s)=\max_{\beta}\log\Big(1+p\|\bh_s\|^2\cos^2(\beta-\alpha)\Big).\label{eq:geo}
\end{align}
Clearly, the maximum rate is achieved if the following function
\begin{align}
f(\beta):=p\|\bh_s\|^2\cos^2(\beta-\alpha)\label{eq:fbetadef}
\end{align}
is maximized.

Moreover, it can be proved by contradiction that the interference
constraint is satisfied with equality, i.e.,
$\bh_{\text{opt}}^H\bS\bh_{\text{opt}}=P_{\th}$. Thus, we have
\begin{align}
p\bh_{\text{opt}}^H\bv\bv^H\bh_{\text{opt}}=p(\bh_0+\sqrt{\epsilon}\sigma\bv)^H\bv\bv^H(\bh_0+\sqrt{\epsilon}\sigma\bv)
=p\big(\|\bh_0\|\cos
\beta+\sqrt{\epsilon}\sigma\big)^2=P_{\th}.\label{eq:inter}
\end{align}
Hence, the interference constraint is transformed into
\begin{align}\label{eq:Pth}
p=\frac{P_{\th}}{\big(\|\bh_0\|\cos\beta+\sqrt{\epsilon}\sigma\big)^2}.
\end{align}
By substituting \eqref{eq:Pth} into \eqref{eq:fbetadef}, we have
\begin{equation}\label{eq:fbeta}
f(\beta)=p\|\bh_s\|^2\cos^2(\beta-\alpha)=\frac{\|\bh_s\|^2P_{\th}\cos^2(\beta-\alpha)}{\big(\|\bh_0\|\cos(\beta)+\sqrt{\epsilon}\sigma\big)^2}.
\end{equation}
Thus, the optimal $\beta_{\text{opt}}$ can be expressed as
\begin{equation}\label{eq:betaopt}
\beta_{\text{opt}}=\arg \max
f(\beta)=\arg\max\frac{\|\bh_s\|^2P_{\th}\cos^2(\beta-\alpha)}{\big(\|\bh_0\|\cos(\beta)+\sqrt{\epsilon}\sigma\big)^2}.
\end{equation}
The problem of \eqref{eq:betaopt} is a single variable optimization
problem. It is easy to observe that the feasible region for $\beta$
is $[\alpha,\pi/2]$. According to the sufficient and necessary
condition for the optimal solution of an optimization problem,
$\beta_{\text{opt}}$ lies either on the border of the region
($\alpha$ or $\pi/2$) or on the point which satisfies $\partial
f(\beta)/\partial \beta=0$. Since
\begin{equation}
\frac{\partial f(\beta)}{\partial
\beta}=\frac{2\|\bh_s\|^2P_{\th}\cos(\beta-\alpha)\Big(\sin\alpha-\sin(\beta-\alpha)\sqrt{\epsilon}\sigma/\|\bh_0\|\Big)}{\|\bh_0\|^2\big(\cos\beta+\sqrt{\epsilon}\sigma/\|\bh_0\|\big)^3},
\end{equation}
we can obtain a locally optimal solution
$\beta_1=\sin^{-1}\Big(\frac{\|\bh_0\|\sin\alpha}{\sqrt{\epsilon}\sigma}\Big)+\alpha$
by solving the equation $\partial f(\beta)/\partial \beta=0$. In the
case when $\frac{\|\bh_0\|\sin\alpha}{\sqrt{\epsilon}\sigma}>1$,
$f(\beta)$ is a non-decreasing function. Hence, the optimal $\beta$
is $\pi/2$, and we define $f(\beta_1)=-\infty$ for this case.
Therefore, the globally optimal solution is
\begin{equation}\label{eq:optbeta}
\beta_{\text{opt}}=\arg\max(f(\alpha),f(\pi/2),f(\beta_1)).
\end{equation}

The optimal power $p_{\text{opt}}$ can be further obtained by
substituting $\beta_{\text{opt}}$ into \eqref{eq:Pth}. According to
the definition of $\beta$ and Lemma \ref{lemma:twodimension}, we
have
\begin{align}\label{eq:optv}
\bv_{\text{opt}}=a_{v}\hat{\bh}+b_{v}\hat{\bh}_{\perp},
\end{align}
where $a_v=\cos(\beta_{\text{opt}})$ and
$b_v=\sin(\beta_{\text{opt}})$. In summary, $\mathbf{SP2}$ can be
solved by Algorithm 1 as described in Table \ref{table:1}.

\subsubsection{Optimal solution to problem
$\mathbf{P3}$}\label{subsection:gloandsub} In the preceding
subsection, we presented the optimal solutions for the two
subproblems. We now turn our attention to the relationship between
problem $\mathbf{P3}$ and the subproblems, and present the complete
algorithm to solve problem $\mathbf{P3}$. Since the convex
optimization problem $\mathbf{P3}$ has two constraints, the optimal
solution can be classified into three cases depending on the
activeness of the constraints: 1) only the transmit power constraint
is active; 2) only the interference constraint is active; and 3)
both constraints are active. Relying on this classification, the
relationship between the solutions of problem $\mathbf{P3}$ and the
two subproblems is described as follows (refer to Appendix
\ref{section:simulprove} for the proof).
\begin{Lemma}\label{lemma:relation}
If the optimal solution $\bS_{1}$ of $\mathbf{SP1}$ satisfies the
constraint of $\mathbf{SP2}$, then $\bS_1$ is the optimal solution
of problem $\mathbf{P3}$. If the optimal solution $\bS_{2}$ of
$\mathbf{SP2}$ satisfies the constraint of $\mathbf{SP1}$, then
$\bS_2$ is the optimal solution of problem $\mathbf{P3}$. Otherwise,
the optimal solution of problem $\mathbf{P3}$ simultaneously
satisfies the transmit power constraint and
$\bh_{\text{opt}}^H\bS\bh_{\text{opt}}\leq P_{\th}$ with equality.
\end{Lemma}

\begin{Remark}To apply Lemma \ref{lemma:relation}, we need to test whether $\bS_1$ and $\bS_2$ satisfy both constraints.
The condition that $\bS_1$ satisfies the interference constraint is
\begin{equation}
P_{\text{int}}\leq
P_{\th},\text{where}~P_{\text{int}}=\max_{\sbh}\bh^H \bS_{1}
\bh,~\text{for}~\|\bh-\bh_0\|^2\leq \epsilon\sigma^2,
\end{equation}
where $P_{\text{int}}$ can be obtained by the method discussed in
Appendix \ref{appdx:lemma}. The condition that $\bS_2$ satisfies the
transmit power constraint is ${\tt{tr}}(\bS_2)\leq \bar{P}$.
\end{Remark}

We next discuss the method for finding the solution in the case
where neither $\bS_1$ nor $\bS_2$ is the optimal solution of problem
$\mathbf{P3}$. Similarly to the method in the preceding subsection,
we solve this case from a geometric perspective. According to Lemma
\ref{lemma:relation}, in the case in which neither $\bS_1$ nor
$\bS_2$ is the feasible solution, the optimal covariance
$\bS_{\text{opt}}$ must satisfy both constraints with equality,
i.e.,
\begin{align}
p_{\text{opt}}=\bar{P},~\text{and}~
p_{\text{opt}}\bh_{\text{opt}}^H\bv_{\text{opt}}\bv_{\text{opt}}^H\bh_{\text{opt}}=P_{\th}.
\end{align}
Combining these two equalities, we have
\begin{equation}
\bar{P}\big(\|\bh_0\|\cos(\beta)+\sqrt{\epsilon}\sigma\big)^2=P_{\th}.
\end{equation}
Thus,
\begin{equation}\label{eq:optbeta2}
\beta_{\text{opt}}=\arccos\Big(\frac{\sqrt{P_{\th}/\bar{P}}-\sqrt{\epsilon}\sigma}{\|\bh_0\|}\Big).
\end{equation}
Based on $\beta_{\text{opt}}$, we can obtain $\bv_{\text{opt}}$ from
\eqref{eq:optv}. We summarize the procedure called Algorithm 2,
which solves the case where both constraints are active for problem
$\mathbf{P3}$, in Table \ref{table:2}. Furthermore, we are now ready
to present the complete algorithm, namely Algorithm 3, to solve
problem $\mathbf{P3}$ in Table \ref{table:3}.

In Algorithm 3, we obtain the optimal solutions to $\mathbf{SP1}$
and $\mathbf{SP2}$ and the optimal solution to the case where both
constraints are active separately. According to Lemma
\ref{lemma:relation}, the final solution obtained in Algorithm 3 is
thus the optimal solution of problem $\mathbf{P3}$.
\begin{theorem}
Algorithm 3 obtains the optimal solution of problem $\mathbf{P3}$.
\end{theorem}

\subsection{The Analytical Method for Problem
$\mathbf{P1}$}\label{subsection:meancov}

In the preceding subsection, the mean feedback problem $\mathbf{P3}$
is solved via a closed-form algorithm. Unlike problem $\mathbf{P3}$,
problem $\mathbf{P1}$ has a non-identity-matrix covariance feedback.
To exploit the closed-form algorithm, we first transform problem
$\mathbf{P1}$ into a problem with the mean feedback form as follows.
\begin{align}
\begin{split}\label{prob:P4}
{\textbf {Equivalent
problem ($\mathbf{P4}$): }} \quad \max_{p,\bar{\sbv}} &\log(1+p\bar{\bh}_s^H \bar{\bv}\bar{\bv}^H \bar{\bh}_s)\\
\text{subject
to}:~&p\|\mathbf{\Delta}^{1/2}\bar{\bv}\|^2\le\bar{P},~p\bar{\bh}^H
\bar{\bv}\bar{\bv}^H\bar{\bh}\leq P_{\th},~
\text{for}~\|\bar{\bh}-\bar{\bh}_0\|^2\leq \epsilon,
\end{split}
\end{align}
where $\bR^{-1}:=\bU^H\mathbf{\Delta}\bU$ obtained by
eigen-decomposing $\bR^{-1}$,
$\bar{\bh}:=\mathbf{\Delta}^{1/2}\bU\bh$,
$\bar{\bh}_0:=\mathbf{\Delta}^{1/2}\bU\bh_0$,
$\bar{\bh}_s:=\mathbf{\Delta}^{1/2}\bU\bh_s$, and
$\bar{\bv}:=\mathbf{\Delta}^{-1/2}\bU\bv$. By substituting these
definitions into \eqref{prob:P4}, it can be observed that the
achieved rates and constraints of both problem $\mathbf{P1}$ and
$\mathbf{P4}$ are equivalent. Thus, the optimal solution of
$\mathbf{P1}$ can be obtained by solving its equivalent problem
$\mathbf{P4}$. Moreover, the optimal beamforming vector
$\bar{\bv}_{\text{opt}}$ of problem $\mathbf{P4}$ can be easily
transformed into the optimal solution $\bv_{\text{opt}}$ for problem
$\mathbf{P1}$ by letting
$\bv_{\text{opt}}=\bU^H\mathbf{\Delta}^{1/2}\bar{\bv}_{\text{opt}}$.
Note that it is not necessary that $\|\bar{\bv}\|=1$ in
\eqref{prob:P4}.

In the preceding subsection, decoupling the multiple constraint
problem into several single constraint subproblems facilitates the
analysis and simplifies the process of solving the problem. For
problem $\mathbf{P4}$, it can also be decoupled into two subproblems
as follows.
\begin{align}
\textbf{Subproblem~3~($\mathbf{SP3}$):}~&\underset{p,\bar{\sbv}}{\max} \log(1+p\bar{\bh}_s^H \bar{\bv}\bar{\bv}^H \bar{\bh}_s)\\
&\text{subject to}:~p\|\mathbf{\Delta}^{1/2}\bar{\bv}\|^2\le\bar{P}.\\
\textbf{Subproblem~4~($\mathbf{SP4}$):}~&\underset{p,\bar{\sbv}}{\max} \log(1+p\bar{\bh}_s^H \bar{\bv}\bar{\bv}^H \bar{\bh}_s)\\
&\text{subject to}:~ p\bar{\bh}^H \bar{\bv}\bar{\bv}^H\bar{\bh}\leq
P_{\th}~ \text{for}~\|\bar{\bh}-\bar{\bh}_0\|^2\leq \epsilon.
\end{align}

It is easy to observe that $\mathbf{SP3}$ is equivalent to
$\mathbf{SP1}$, and the optimal transmit covariance matrix of
$\mathbf{SP3}$ can be obtained in the same way as that for
$\mathbf{SP1}$. Moreover, $\mathbf{SP4}$ is the same as
$\mathbf{SP2}$, and thus it can be solved by Algorithm 1 discussed
in Subsection \ref{subsection:subsolution}.

The relationship between problem $\mathbf{P4}$ and subproblems
$\mathbf{SP3}$ and $\mathbf{SP4}$ is similar to the one between
$\mathbf{P3}$ and corresponding subproblems as depicted in Lemma
\ref{lemma:relation}, i.e., if either optimal solution of
$\mathbf{SP3}$ or $\mathbf{SP4}$ satisfies both constraints, then it
is the globally optimal solution; otherwise, the optimal solution
satisfies both constraints with equalities. We hereafter need to
consider only the case in which the solutions of both subproblems
are not feasible for problem $\mathbf{P4}$. For this case, the two
equality constraints can be written as follows.
\begin{align}\label{eq:twoconstriant}
\|\mathbf{\Delta}^{1/2}\bar{\bv}\|=1,~{\text {and}}~
\max\big(\bar{\bh}^H\bar{\bv}\bar{\bv}^H\bar{\bh}\big)=\frac{P_{\th}}{\bar{P}},~\text{for}~
\|\bar{\bh}-\bar{\bh}_0\|^2\le
 \epsilon.
\end{align}
Assume that the angle between $\bar{\bh}_0$ and $\bar{\bv}$ is
$\bar{\beta}$, and that $\bar{p}:=\|\bar{\bv}\|$. Similar to Lemma
\ref{lemma:twodimension}, the optimal $\bar{\bv}$ lies in a plane
spanned by $\hat{\bar{\bh}}$ and $\hat{\bar{\bh}}_{\perp}$, where
$\hat{\bar{\bh}}=\bar{\bh}_0/\|\bar{\bh}_0\|$,
$\hat{\bar{\bh}}_{\perp}=\bar{\bh}_{\perp}/\|\bar{\bh}_{\perp}\|$,
and
$\bar{\bh}_{\perp}=\bar{\bh}_s-(\hat{\bar{\bh}}^H\bar{\bh}_s)\hat{\bar{\bh}}$.
Thus, if we can determine $\bar{\beta}$ and $\bar{p}$ from
\eqref{eq:twoconstriant}, then the optimal $\bar{\bv}$ can be
identified by
\begin{equation}\label{eq:vbardefine}
\bar{\bv}=\bar{p}\big(\cos(\bar{\beta})\hat{\bar{\bh}}+\sin(\bar{\beta})\hat{\bar{\bh}}_{\perp}\big).
\end{equation}

Based on the new variables $\bar{\beta}$ and $\bar{p}$, the
constraints \eqref{eq:twoconstriant} can be transformed as follows.
\begin{align}
\bar{p}\Big\|\mathbf{\Delta}^{1/2}\big(\cos(\bar{\beta})\hat{\bar{\bh}}+\sin(\bar{\beta})\hat{\bar{\bh}}_{\perp}\big)\Big\|&=1,\label{eq:pbarcons}\\
\text{and},~\bar{p}\big(\cos(\bar{\beta})\|\bar{\bh}_0\|+\sqrt{\epsilon}\big)&=\sqrt{\frac{P_{\th}}{\bar{P}}}.\label{eq:Pthcons}
\end{align}
According to \eqref{eq:pbarcons}, we have
\begin{align}
\bar{p}=\frac{1}{\Big\|\mathbf{\Delta}^{1/2}\big(\cos(\bar{\beta})\hat{\bar{\bh}}+\sin(\bar{\beta})\hat{\bar{\bh}}_{\perp}\big)\Big\|}.\label{eq:t}
\end{align}
Substituting \eqref{eq:t} into \eqref{eq:Pthcons}, we have
\begin{align}
\sqrt{\frac{P_{\th}}{\bar{P}}}\Big\|\mathbf{\Delta}^{1/2}\big(\cos(\bar{\beta})\hat{\bar{\bh}}+\sin(\bar{\beta})\hat{\bar{\bh}}_{\perp}\big)\Big\|=\cos(\bar{\beta})\|\bar{\bh}_0\|+\sqrt{\epsilon}.\label{eq:solvebeta}
\end{align}
Hence, the optimal $\bar{\beta}$ can be obtained by solving
\eqref{eq:solvebeta}, and $\bar{\bv}_{\text{opt}}$ can be obtained
by substituting $\bar{\beta}$ into \eqref{eq:vbardefine}. In
summary, the procedure to solve the case in which both constraints
are active is listed as Algorithm 4 in Table \ref{table:4}.
Moreover, we are now ready to present the complete algorithm, namely
Algorithm 5, for solving problem $\mathbf{P1}$ in Table
\ref{table:5}.

In Algorithm 5, we obtain the optimal solutions to $\mathbf{SP3}$
and $\mathbf{SP4}$ and the optimal solution to the case where both
constraints are active separately. According to Lemma
\ref{lemma:relation}, the final result obtained in Algorithm 5 is
thus the optimal solution of problem $\mathbf{P1}$.
\begin{theorem}
Algorithm 5 achieves the optimal solution of problem $\mathbf{P1}$.
\end{theorem}

\begin{Remark}
The complexity of the interior point algorithm for the SOCP problem
\eqref{eq:SDPreal} is
$\mathcal{O}(N^{3.5}\log(\frac{1}{\varepsilon}))$, where
$\varepsilon$ denotes the error tolerance. For Algorithm 5, a
maximum of $\mathcal{O}(\log(\frac{1}{\varepsilon}))$ operations is
needed to solve \eqref{eq:solvebeta}, and the complexity for each
operation is $\mathcal{O}(\log(N^2))$. Hence, the computation
complexity required for Algorithm 5 is
$\mathcal{O}(N^2\log(\frac{1}{\varepsilon}))$, which is much less
than that of the interior point algorithm.
\end{Remark}

\section{Simulations}\label{section:simulation}

Computer simulations are provided in this section to evaluate the
performance of the proposed algorithms. In the simulations, it is
assumed that the entries of the channel vectors $\bh_s$ and $\bh_0$
are modeled as independent circularly symmetric complex Gaussian
random variables with zero mean and unit variance. Moreover, we
denote by $l_1$ the distance between the SU-Tx and the SU-Rx, and by
$l_2$ the distance between the SU-Tx and the PU. It is assumed that
the same path loss model is used to describe the transmissions from
the SU-Tx to the SU-Rx and to the PU, and the path loss exponent is
chosen to be $4$. The noise power is chosen to be $1$, and the
transmit power and interference power are defined in dB relative to
the noise power. For all cases, we choose $P_{\th} = 0$ dB.

\subsection{Comparison of the Analytical Solution and the Solution Obtained by the SOCP Algorithm}

\begin{figure}
      \centering
      \includegraphics[width = 150mm]{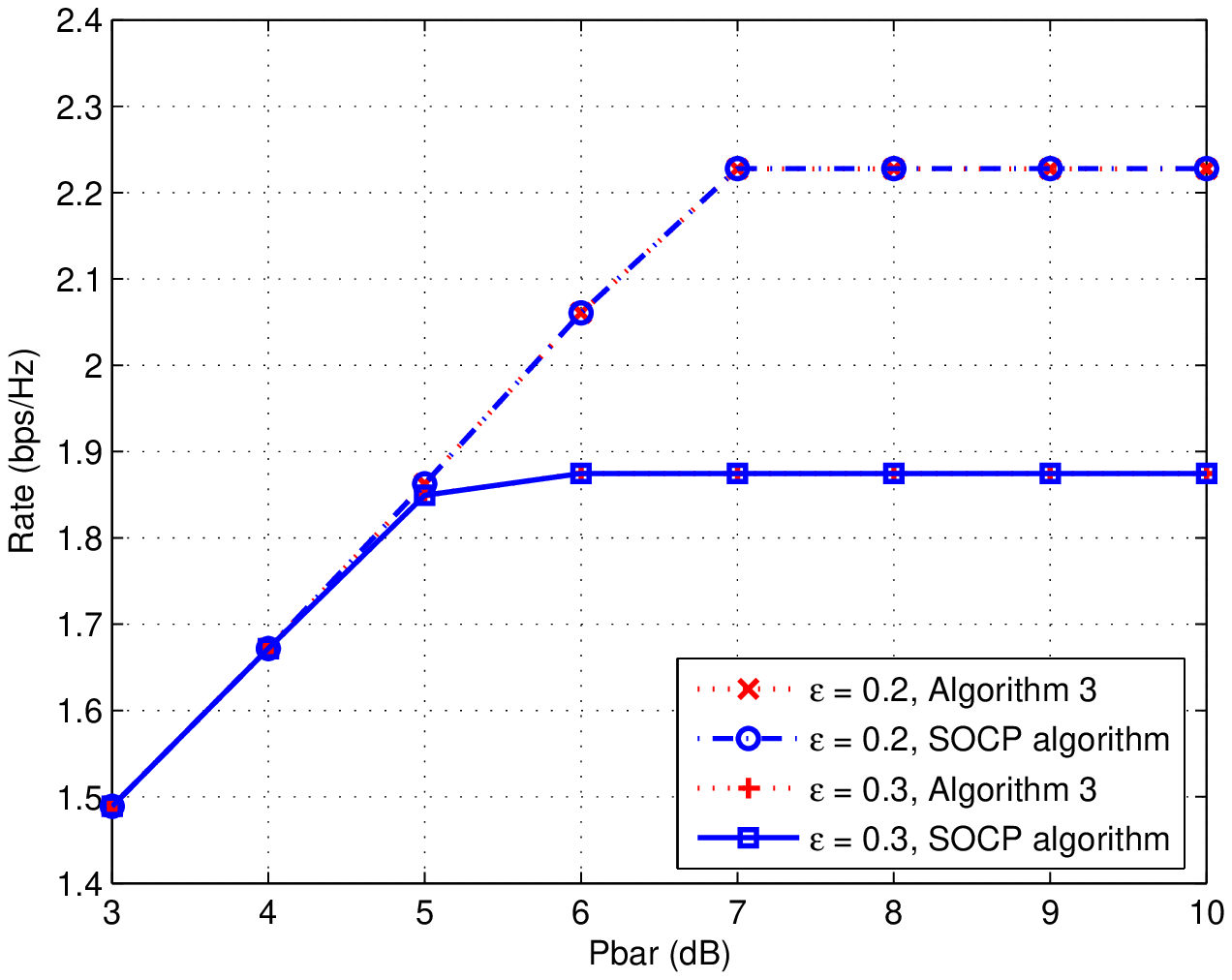}
     \caption{Comparison of the results obtained by the SOCP algorithm and Algorithm 3.}
     \label{fig:compare}
\end{figure}

\begin{figure}
      \centering
      \includegraphics[width = 150mm]{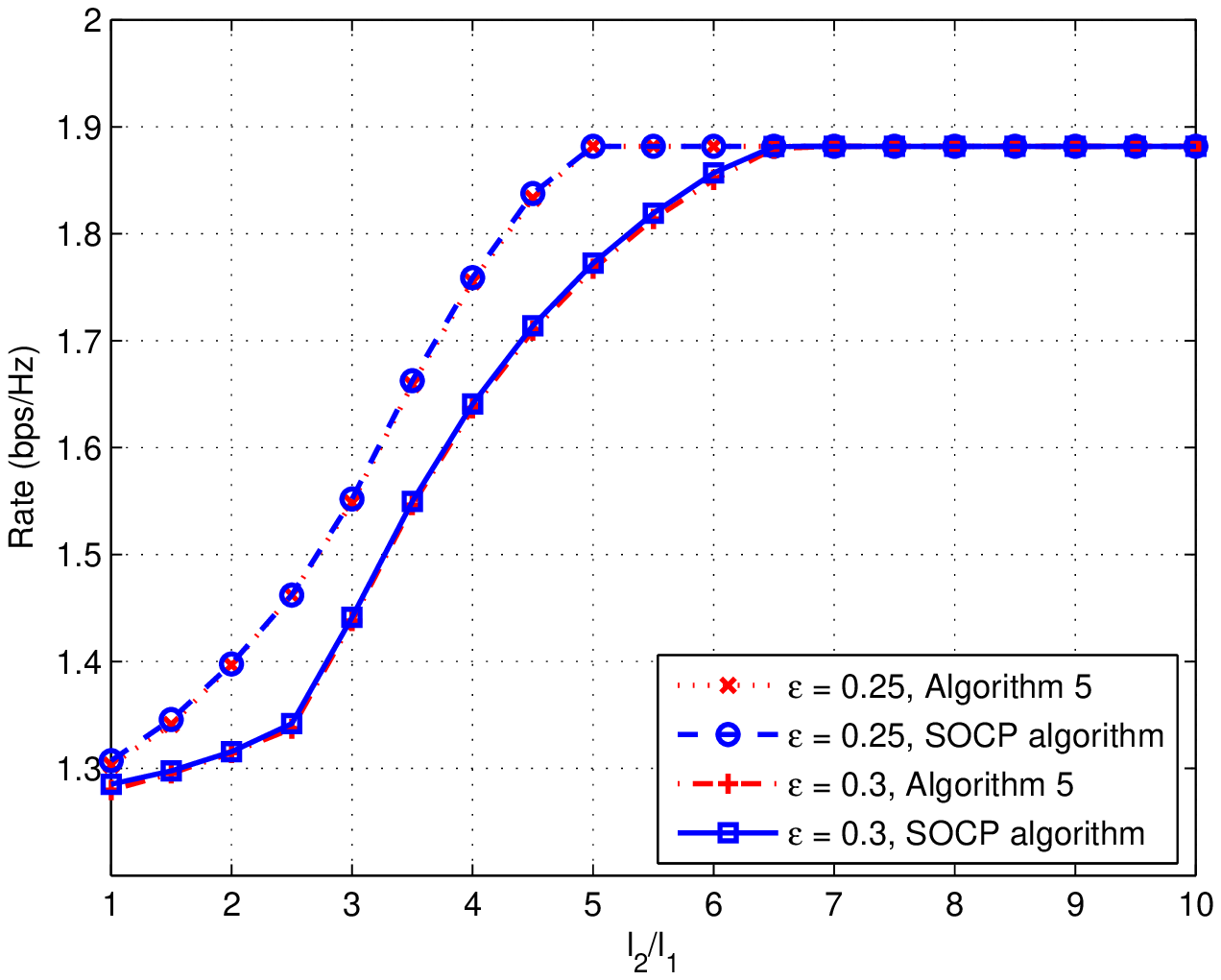}
     \caption{Comparison of the results obtained by the SOCP algorithm and Algorithm 5.}
     \label{fig:compare1}
\end{figure}

In this simulation, we compare the two results obtained by a
standard SOCP algorithm (SeDuMi) and Algorithm 3. We consider the
system with $N=3$, $l_2/l_1=2$, and $\bar{P}$ ranging from 3 dB to
10 dB. In Fig. \ref{fig:compare}, we can see that the results
obtained by different algorithms coincide. This is because both
algorithms determine the optimal solution. Compared with the SOCP
algorithm solution, Algorithm 3 obtains the solution directly, and
thus it has lower complexity. In Fig. \ref{fig:compare1}, we compare
the two results obtained by SeDuMi and Algorithm 5. We consider the
system with $N=3$, $\bar{P}$ =  5 dB, and $l_2/l_1$ ranging from 1
to 10. The covariance matrix $\bR$ is generated by $\bR_1^H\bR_1$,
where each element of $\bR_1$ follows Gaussian distribution with
zero mean and unit variance. From Fig. \ref{fig:compare1}, we can
see that the results obtained by the two algorithms coincide again.
Moreover, we note that the achievable rate with $\epsilon=0.2$ is
always greater than or equal to the rate with $\epsilon=0.3$, since
a larger $\epsilon$ corresponds to the stricter constraints.

\subsection{Effectiveness of the Interference Constraint}
\begin{figure}[t]
      \centering
     \includegraphics[width = 150mm]{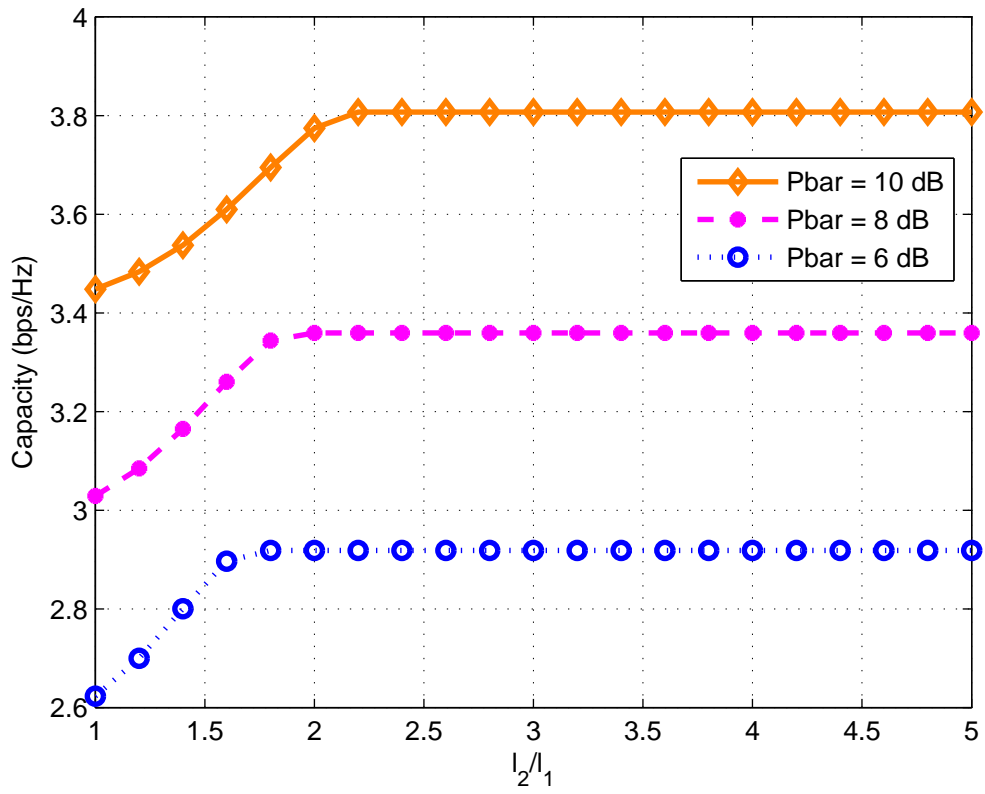}
     \caption{Effect of $l_{2}/l_{1}$ on the achievable rate of the CR network ($\epsilon=1$, $N=3$).}
     \label{fig:l2inc}
\end{figure}

In this simulation, we apply Algorithm 3 to solve problem
$\mathbf{P3}$. In Fig. \ref{fig:l2inc}, we depict the achievable
rate versus the ratio $l_{2}/l_{1}$ under different transmit power
constraints. The increase of the ratio $l_{2}/l_{1}$ corresponds the
decrease of the interference power constraint. As shown in Fig.
\ref{fig:l2inc}, with an increase of $l_{2}/l_{1}$, the achievable
rate increases due to the lower interference constraint. Until the
ratio $l_2/l_1$ reaches a certain value, the achievable rate remains
unchanged, since the transmit power constraint dominates the result,
and the interference constraint becomes inactive.


\subsection{The Activeness of the Constraints}
\begin{figure}[t]
      \centering
     \includegraphics[width = 150mm]{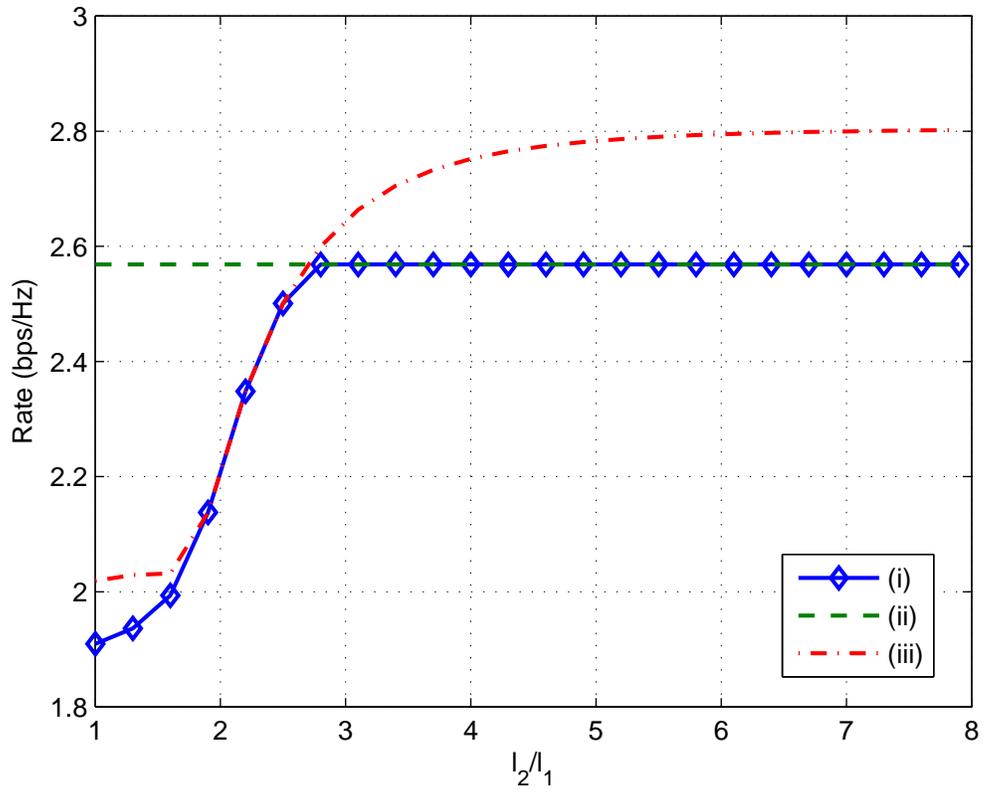}
     \caption{Comparison of the rate under different constraints of problem $\mathbf{P1}$. (i) the maximal rate subject to interference constraint and transmit power constraint simultaneously; (ii) the maximal rate subject to a single transmit power constraint;
     (iii) the maximal rate subject to a single interference constraint.}
     \label{fig:activity}
\end{figure}

In this simulation, we compare the achieved rates of problem
$\mathbf{P1}$ with a single transmit power constraint, a single
interference constraint and both constraints. Here, we choose $N=3$,
$\epsilon = 0.2$, and generate $\bR$ in the same way as in the first
simulation example. Fig. \ref{fig:activity} plots three achievable
rates for different constraints, respectively. It can be observed
from Fig. \ref{fig:activity} that the rate under two constraints is
always less than or equal to the rate under a single constraint.
Obviously, this is due to the fact that extra constraints reduce the
degree of freedom of the transmitter.

\section{Conclusions}\label{section:conclusion}

In this paper, the robust cognitive beamforming design problem has
been investigated, for the SU MISO communication system in which
only partial CSI of the link from the SU-Tx to the PU is available
at the SU-Tx. The problem can be formulated as an SIP optimization
problem. Two approaches have been proposed to obtain the optimal
solution of the problem; one approach is based on a standard
interior point algorithm, while the other approach solves the
problem analytically. Simulation examples have been used to present
a comparison of the two approaches as well as to study the
effectiveness and activeness of imposed constraints.

This work initiates research in robust design of cognitive radios.
We are currently extending these methods to the more general case
with multiple receive antennas and multiple PUs. Other interesting
extensions include more practical scenarios, such as the case in
which the SU channel information is also partially known at the
SU-Tx.

\def\appref#1{Appendix~\ref{#1}}
\appendix
\renewcommand{\thesubsection}{\Alph{subsection}}
\makeatletter
\renewcommand{\subsection}{%
\@startsection {subsection}{2}{\z@ }{2.0ex plus .5ex minus .2ex}%
{-1.0ex plus .2ex}{\it }} \makeatother

\subsection{Proof of Lemma \ref{lemma:1D}}\label{section:prof1D}

Problem $\mathbf{P1}$ involves infinitely many constraints. Denote
the set of active constraints by $\mathcal{C}$, the cardinality of
the set $\mathcal{C}$ by $K$, and the channel response related to
the $k$th element of the set $\mathcal{C}$ by $\bh_k$. According to
the Karush-Kuhn-Tucker (KKT) conditions for $\mathbf{P1}$, we have:
\begin{align}
\bh_s(1+\bh_s^H\bS\bh_s)^{-1}\bh_s^H+\bPhi&=\lambda\bI+\sum_{i=1}^{K}\mu_i\bh_i\bh_i^H,\label{kkt:1}\\
{\tt{tr}}(\bPhi\bS)&=0\label{kkt:4},
\end{align}
where $\bPhi$ is the dual variable associated with the constraint
$\bS\geq 0$, and $\lambda$ and $\mu_i$ are the dual variables
associated with the transmit power constraint and the interference
constraint, respectively. First, we assume that $\lambda\neq 0$, and
thus the rank of the right hand side of \eqref{kkt:1} is $N$. Since
the first term on the left hand side of \eqref{kkt:1} has rank one,
we have
\begin{equation}
{{\tt{Rank}}}(\bPhi)\geq N-1.\label{eq:greN}
\end{equation}
Moreover, since $\bS\geq 0$ and $\bPhi\geq0$, from \eqref{kkt:4} we
have ${\tt{tr}}(\bPhi\bS)={\tt{tr}}(\bU^H\bLambda
\bU\bS)={\tt{tr}}(\bLambda \bU\bS
\bU^H)={\tt{tr}}(\bLambda\tilde{\bS})=0$, where $\bU^H\bLambda \bU$
is the eigenvalue decomposition of matrix $\bPhi$, and
$\tilde{\bS}:=\bU\bS \bU^H$. By applying eigenvalue decomposition to
$\tilde{\bS}$, we have $\tilde{\bS}:=\sum_i\tau_i\bs_i\bs_i^H$,
where $\tau_i$ is the $i$th eigenvalue and $\bs_i$ is the
corresponding eigenvector. We next show
${\tt{Rank}}(\bS)+{\tt{Rank}}(\bPhi)\leq N$ by contradiction.
Suppose that ${{\tt{Rank}}}(\bS)+{{\tt{Rank}}}(\bPhi)> N$. Then,
there exists an index $j$ such that the $j$th element of $\bs_i$ and
the $j$th diagonal element of $\bLambda$ are non-zero
simultaneously. Thus, it is impossible that the equation
${\tt{tr}}(\bLambda\tilde{\bS})=0$ holds. It follows that
${{\tt{Rank}}}(\bS)+{{\tt{Rank}}}(\bPhi)\leq N$. Combining this with
\eqref{eq:greN}, we have ${{\tt{Rank}}}(\bS)\leq 1$.

Second, we assume that $\lambda=0$ in \eqref{kkt:1}. In this case,
$\bS$ must lie in the space spanned by $\bh_i$, $i=1,\cdots,K$. Let
the dimensionality of the space be $M$. Therefore, we can restrict
${{\tt{Rank}}}(\bPhi)\le M$. Thus, the reminder of the proof is the
same as that of the case $\lambda\neq0$, and the proof is
complete.\hfill $\blacksquare$

\subsection{Proof of Lemma \ref{lemma:cond}
}\label{section:profcond} First, we consider the sufficiency part of
this lemma. We assume that there exists a covariance matrix
$\bS_{\text{opt}}$ and an $\bh_{\text{opt}}$ that satisfy the
conditions \eqref{prob:finite} and \eqref{eq:sufficienth}
simultaneously. Since $\bS_{\text{opt}}$ satisfies both the transmit
power constraint and the interference constraint, $\bS_{\text{opt}}$
is a feasible solution for problem $\mathbf{P1}$. Moreover, if we
assume that there exists another solution $\bS_{s}$, which results
in a larger achievable rate for the SU link, then a contradiction
will be derived. Without loss of generality, we assume that the
constraint set, which consists of all the active interference
constraints for $\bS_{s}$, is denoted by $\mathcal{T}$. We divide
the set $\mathcal{T}$ into two types: one type is
$\bh_{\text{opt}}\in \mathcal{T}$, and the other type is
$\bh_{\text{opt}}\notin \mathcal{T}$.

Assume that $C_{s}$ and $C_{\text{opt}}$ are the achievable rates
for the covariance matrices $\bS_{s}$ and $\bS_{\text{opt}}$,
respectively. In the case of $\bh_{\text{opt}}\in \mathcal{T}$, we
have $C_{s}\leq C_{\text{opt}}$, since $C_{\text{opt}}$ is obtained
with fewer constraints. Since problem $\mathbf{P1}$ is a convex
optimization problem that has a unique optimal solution,
$\bS_{\text{opt}}$ is indeed the optimal solution. In the case of
$\bh_{\text{opt}}\notin \mathcal{T}$, we can observe that
$\bS_{\text{opt}}$ satisfies the constraints in $\mathcal{T}$, and
$\bS_{s}$ satisfies the constraint $\bh_{\text{opt}}$. According to
the lemma in \cite{lan07:jsac}, this case does not exist.

We next proceed to prove the necessity part. Suppose that
$\bS_{\text{opt}}$ is the optimal solution of problem $\mathbf{P1}$.
According to Lemma \ref{lemma:1D}, we have
$\bS_{\text{opt}}=p_{\text{opt}}\bv_{\text{opt}}\bv_{\text{opt}}^H$.
Thus, problem $\mathbf{P1}$ is equivalent to
\begin{align}\label{prob:equ}
\begin{split}
&\max_{\sbS\geq0}~\log(1+\bh_s^H \bS \bh_s)\\
\text{subject to}:~&{\tt{tr}}(\bS)\leq p_{\text{opt}},~\bh^H \bS
\bh\leq P_{\th},~\text{for}~(\bh-\bh_0)^H\bR^{-1}(\bh-\bh_0)\leq
\epsilon.
\end{split}
\end{align}

According to Lemma \ref{lemma:par}, there is a unique
\begin{equation}
\bh_{\text{opt}}=\bh_0+\sqrt{\frac{\epsilon}{\bv_{\text{opt}}^H\bR\bv_{\text{opt}}}}\alpha\bR
\bv_{\text{opt}},
\end{equation}
which is the optimal solution of
$\max_{\sbh\in\mathcal{H}(\epsilon)}\bh^H\bS\bh\le P_{\th}$. Thus,
for problem \eqref{prob:equ}, only ${\tt{tr}}(\bS)\leq
p_{\text{opt}}$ and $\bh_{\text{opt}}^H \bS \bh_{\text{opt}}\leq
P_{\th}$ are active constraints. Thus, it is obvious that problem
\eqref{prob:equ} and problem \eqref{prob:finite} have the same
optimal solution. Hence, the proof is complete. \hfill
$\blacksquare$

%
%

\subsection{Proof of Lemma \ref{lemma:twodimension}
}\label{section:prov2D}

The proof of Lemma \ref{lemma:twodimension} is divided into two
parts. The first part is to prove that $\bv_{\text{opt}}$ is in the
form of $\alpha_{v}\hat{\bh}+\beta_{v}\hat{\bh}_{\perp}$, where
$\alpha_v\in\mathbb{C}$ and $\beta_v\in\mathbb{C}$. The second part
is to prove $\alpha_v\in\mathbb{R}$ and $\beta_v\in\mathbb{R}$. In
the following proof, we assume that $\alpha_{k}\in\mathbb{C}$ are
some proper complex scalars.

According to Lemma \ref{lemma:cond}, and Theorem 2 in
\cite{Liang:jstsp}, we have
\begin{equation}\label{eq:vexp}
\bv_{\text{opt}}=\alpha_1\bh_{\text{opt}}+\alpha_2\bh_s.
\end{equation}
According to Lemma \ref{lemma:par}, we have
\begin{align}
\bh_{\text{opt}}=\bh_0+\alpha_3\bv_{\text{opt}}=\bh_0+\alpha_3\big(\alpha_1\bh_{\text{opt}}+\alpha_2\bh_s\big)=\bh_0+\alpha_1\alpha_3\bh_{\text{opt}}+\alpha_2\alpha_3\bh_s.\label{eq:provespace1}
\end{align}
According to \eqref{eq:provespace1}, it can be observed that
$\bh_{\text{opt}}$ can be expressed by the linear combination of
$\bh_0$ and $\bh_s$, where the coefficients are complex. Combining
this with \eqref{eq:vexp}, we have
$\bv_{\text{opt}}=\alpha_4\bh_0+\alpha_5\bh_s,$ where
$\alpha_4\in\mathbb{C}$ and $\alpha_5\in\mathbb{C}$. Moreover, since
both $\bh_0$ and $\bh_s$ can be expressed as a linear combination of
$\hat{\bh}$ and $\hat{\bh}_{\perp}$, we have
$\bv_{\text{opt}}=\alpha_{v}\hat{\bh}+\beta_{v}\hat{\bh}_{\perp}.$
Since rotating $\bv_{\text{opt}}$ does not affect the final result,
we can assume $\alpha_v\in\mathbb{R}$.

We next prove that $\beta_v\in \mathbb{R}$ by contradiction. At
first, we assume that $\beta_v=a+jb\notin\mathbb{R}$. Then we can
find an equivalent $\hat{\beta}_v=\sqrt{a^2+b^2}\in \mathbb{R}$
which is a better solution of problem $\mathbf{P1}$ than $\beta_v$.
Assume that
$\hat{\bv}_{\text{opt}}=\alpha_{v}\hat{\bh}+\hat{\beta}_{v}\hat{\bh}_{\perp}$.
It is clear that $\|\hat{\bv}_{\text{opt}}\|=\|\bv_{\text{opt}}\|$,
and the interference caused by $\hat{\bv}_{\text{opt}}$ is
\begin{align}
p\bh_{\text{opt}}^H\hat{\bv}_{\text{opt}}\hat{\bv}_{\text{opt}}^H\bh_{\text{opt}}
=&p(\bh_0+\sqrt{\frac{\epsilon}{\hat{\bv}_{\text{opt}}^H\bR\hat{\bv}_{\text{opt}}}}\alpha\bR \hat{\bv}_{\text{opt}})^H\hat{\bv}_{\text{opt}}\hat{\bv}_{\text{opt}}^H(\bh_0+\sqrt{\frac{\epsilon}{\hat{\bv}_{\text{opt}}^H\bR\hat{\bv}_{\text{opt}}}}\alpha \bR\hat{\bv}_{\text{opt}})\\
=&p\big(\alpha_{v}\|\bh_0\|+\sqrt{\frac{\epsilon}{\hat{\bv}_{\text{opt}}^H\bR\hat{\bv}_{\text{opt}}}}\alpha^H\hat{\bv}_{\text{opt}}^H\bR\hat{\bv}_{\text{opt}}\big)^2,
\end{align}
which is equal to that of $\bv_{\text{opt}}$. However, the
corresponding objective function with $\hat{\bv}_{\text{opt}}$ is
\begin{align}
\log(1+p\bh_s^H\hat{\bv}_{\text{opt}}\hat{\bv}_{\text{opt}}^H\bh_s)
=&\log(1+p(a_{h_s}\hat{\bh}+b_{h_s}\hat{\bh}_{\perp})^H(\alpha_{v}\hat{\bh}+\hat{\beta}_{v}\hat{\bh}_{\perp})(\alpha_{v}\hat{\bh}+\hat{\beta}_{v}\hat{\bh}_{\perp})^H(a_{h_s}\hat{\bh}+b_{h_s}\hat{\bh}_{\perp}))\notag\\
=&\log(1+p(a_{h_s}\alpha_v+b_{h_s}\hat{\beta}_{v})(a_{h_s}\alpha_v+b_{h_s}\hat{\beta}_{v}^H)),\label{eq:proofobj}
\end{align}
and the objective value with $\bv_{\text{opt}}$ is
\begin{align}
\log(1+p\bh_s^H\bv_{\text{opt}}\bv_{\text{opt}}^H\bh_s)
=&\log(1+p(a_{h_s}\hat{\bh}+b_{h_s}\hat{\bh}_{\perp})^H(\alpha_{v}\hat{\bh}+\beta_{v}\hat{\bh}_{\perp})(\alpha_{v}\hat{\bh}+\beta_{v}\hat{\bh}_{\perp})^H(a_{h_s}\hat{\bh}+b_{h_s}\hat{\bh}_{\perp}))\notag\\
=&\log(1+p(a_{h_s}\alpha_v+b_{h_s}\beta_{v})(a_{h_s}\alpha_v+b_{h_s}\beta_{v}^H)).\label{eq:proofobj1}
\end{align}

According to \eqref{eq:proofobj} and \eqref{eq:proofobj1}, we can
conclude that $\hat{\bv}_{\text{opt}}$ is a better solution. The
proof follows.

\subsection{Lemma \ref{lemma:par} and its proof}\label{appdx:lemma}
\begin{Lemma}\label{lemma:par}
For the problem
\begin{align}
\max_{\sbh}~ p\bh^H\bv\bv^H\bh, ~\text{subject
to:}~(\bh-\bh_0)^H\bR^{-1}(\bh-\bh_0)\leq \epsilon,
\end{align}
where $p$, $\bv$, and $\bh_0$ are constant, the optimal solution is
\begin{align}
\bh_{\text{max}}=\bh_0+\sqrt{\frac{\epsilon}{\bv^H\bR\bv}}\alpha\bR\bv,\text{where}~\alpha=\bv^H\bh_0/|\bv^H\bh_0|.
\end{align}
\end{Lemma}

\begin{proof}
The objective function $p\bh^H\bv\bv^H\bh$ is a convex function. The
duality gap for a convex maximization problem is zero. The
Lagrangian function is
\begin{align}
L(\bh,\lambda)=p\bh^H\bv\bv^H\bh-\lambda\Big((\bh-\bh_0)^H\bR^{-1}(\bh-\bh_0)-\epsilon\Big),
\end{align}
where $\lambda$ is the Lagrange multiplier. According to the KKT
condition, we have $\frac{\partial L}{\partial
\bh}=2p\bv\bv^H\bh-2\lambda \bR^{-1}(\bh-\bh_0)=0.$ Thus,
\begin{align}\label{eq:proofpar}
p(\bv^H\bh)\bv=\lambda\bR^{-1}(\bh-\bh_0).
\end{align}
We have $\bh_{\max}=\bh_0+b\alpha\bR\bv$, where $b\in \mathbb{R}$,
$\alpha\in\mathbb{C}$, and $|\alpha|=1$. Since
$(\bh-\bh_0)^H\bR^{-1}(\bh-\bh_0)= \epsilon$, we have
$b=\sqrt{\epsilon}/\sqrt{\bv^H\bR^H\bv}$. Moreover, by observing
\eqref{eq:proofpar}, we have
$\alpha=t\bv^H\bh=t\bv^H(\bh_0+b\alpha\bR\bv)=t\bv^H\bh_0+tb\alpha\bv^H\bR\bv,$
where $t$ is a real scalar such that $|t\bv^H\bh|=1$. Thus, we have
$\bv^H\bh_0/|\bv^H\bh_0|=\alpha$. The proof follows immediately.
\end{proof}

\subsection{Proof of Lemma
\ref{lemma:subcond}}\label{section:profsubcond} Similar to the proof
of Lemma \ref{lemma:cond}, we can show that the problem
\begin{align}\label{eq:Sopt}
\bS_{\text{opt}}=\arg\max_{\sbS,p}\log(1+\bh_s^H \bS \bh_s)~
\text{subject to}:~\bh_{\text{opt}}^H \bS \bh_{\text{opt}}\leq
P_{\th},
\end{align}
where $\bh_{\text{opt}}=\arg\max_{\sbh}\bh^H \bS_{\text{opt}}
\bh,~\text{for}~(\bh-\bh_0)^H\bR^{-1}(\bh-\bh_0)\leq \epsilon$, is
equivalent to $\mathbf{SP2}$.

Since $\bS_{\text{opt}}$ is a rank-1 matrix, according to Lemma
\ref{lemma:par}, we have
$\bh_{\text{opt}}=\bh_0+\sqrt{\epsilon}\sigma\bv.$ Combining this
with \eqref{eq:Sopt}, we have
$\bS_{\text{opt}}=\arg\max_{\bS,p}\log(1+\bh_s^H \bS \bh_s)~
\text{s.t.}:~(\bh_0+\sqrt{\epsilon}\sigma\bv)^H \bS
(\bh_0+\sqrt{\epsilon}\sigma\bv)\leq P_{\th}$, which is equivalent
to \eqref{prob:finitesub}. The proof is complete. \hfill
$\blacksquare$.
\subsection{Proof of Lemma
\ref{lemma:relation}}\label{section:simulprove}

Assume that $\bS_{\text{opt}}$ is the optimal solution for problem
$\mathbf{P3}$. If $\bS_1$ satisfies the interference constraint,
then $\bS_1$ is a feasible solution for problem $\mathbf{P3}$. The
optimal rate achieved by $\bS_{\text{opt}}$ cannot be larger than
that of $\bS_1$, since the constraint of $\mathbf{SP1}$ is a subset
of problem $\mathbf{P3}$. Similarly, we can prove the second part of
the Lemma. We now focus on the third part of this lemma. For problem
$\mathbf{P3}$, at least one of ${\tt{tr}}(\bS)\leq \bar{P}$ and
$\bh_{\text{opt}}^H\bS\bh_{\text{opt}}\leq P_{\th}$ is an active
constraint, since if neither of them is active, we can always find
an $\epsilon$ such that $\bS_{\text{opt}}+\epsilon\bI$ is a feasible
and better solution. Moreover, if only ${\tt{tr}}(\bS)\leq \bar{P}$
is active, then $\bS_1$ is the optimal solution, which contradicts
with $\bh_{\text{opt}}^H\bS_1\bh_{\text{opt}}\geq P_{\th}$.
Similarly, it is impossible that only
$\bh_{\text{opt}}^H\bS\bh_{\text{opt}}\leq P_{\th}$ is active.
Therefore, both constraints are active constraints. \hfill
$\blacksquare$

{\linespread{1.5}

}
\newpage

{\Large
\begin{table}
\caption{The algorithm for SP2.}
\begin{center}
\begin{tabular}{l}
  \hline
  Algorithm 1\\
  \hline
   1. Compute $\beta_{\text{opt}}$ through \eqref{eq:optbeta},\ \ \ \ \ \ \ \ \ \ \ \ \ \ \ \ \ \ \ \ \ \ \ \ \ \\
   2. Compute $p_{\text{opt}}$ according to \eqref{eq:Pth},\\
   3. Compute $\bv_{\text{opt}}$ according to \eqref{eq:optv},\\
   4.
   $\bS_{\text{opt}}=p_{\text{opt}}\bv_{\text{opt}}\bv_{\text{opt}}^H$.\\
  \hline
\end{tabular}\label{table:1}
\end{center}
\end{table}

\begin{table}
\caption{The algorithm for problem $\mathbf{P3}$ in the case where
two constraints are satisfied simultaneously.}
\begin{center}
\begin{tabular}{l}
  \hline
  Algorithm 2\\
  \hline
   1. Compute $\beta_{\text{opt}}$ through \eqref{eq:optbeta2},\ \ \ \ \ \ \ \ \ \ \ \ \ \ \ \ \ \ \ \ \ \ \ \ \ \ \\
   2. Based on \eqref{eq:optv}, compute $\bv_{\text{opt}}$,\\
   3.
   $\bS_{\text{opt}}=\bar{P}\bv_{\text{opt}}\bv_{\text{opt}}^H$.\\
  \hline
\end{tabular}\label{table:2}
\end{center}
\end{table}

\begin{table}
\caption{The complete algorithm for problem $\mathbf{P3}$.}
\begin{center}
\begin{tabular}{l}
  \hline
  Algorithm 3\\
  \hline
   1. Compute the optimal solution $\bS_1=\bar{P}\bh_s\bh_s^H/\|\bh_s\|^2$ for $\mathbf{SP1}$,\\
   2. Compute the optimal solution $\bS_2$ for $\mathbf{SP2}$ via Algorithm 1,\\
   3. If $\bS_1$ satisfies the interference constraint, then $\bS_1$ is the optimal
    solution,\\
   4.
   Elsif $\bS_2$ satisfies the transmit power constraint, then $\bS_2$ is the optimal solution,\\
   5. Otherwise compute the optimal solution via Algorithm 2.\\
  \hline
\end{tabular}\label{table:3}
\end{center}
\end{table}

\begin{table}
\caption{The algorithm for problem $\mathbf{P4}$ in the case where
two constraints are satisfied simultaneously.}
\begin{center}
\begin{tabular}{l}
  \hline
  Algorithm 4\\
  \hline
   1. Compute $\bar{\beta}$ via \eqref{eq:solvebeta}, and compute $\bar{\bv}$ via \eqref{eq:vbardefine},\\
   2. Based on the relationship between $\bar{\bv}$ and $\bv$, compute $\bv_{\text{opt}}$,\ \ \ \ \ \ \ \ \ \ \ \ \ \ \ \ \ \ \ \ \ \ \ \ \ \ \ \ \\
   3.
   $\bS_{\text{opt}}=\bar{P}\bv_{\text{opt}}\bv_{\text{opt}}^H$.\\
  \hline
\end{tabular}\label{table:4}
\end{center}
\end{table}

\begin{table}
\caption{The complete algorithm for problem $\mathbf{P1}$.}
\begin{center}
\begin{tabular}{l}
  \hline
  Algorithm 5\\
  \hline
   1. Compute the optimal solution $\bS_3=\bar{P}\bh_s\bh_s^H/\|\bh_s\|^2$ for $\mathbf{SP3}$,\\
   2. Compute the optimal solution $\bS_4$ for $\mathbf{SP4}$ via Algorithm 4,\\
   3. If $\bS_3$ satisfies the interference constraint, then $\bS_3$ is the optimal solution,\\
   4.
   Elsif $\bS_4$ satisfies the transmit power constraint, then $\bS_4$ is the optimal solution,\\
   5. Otherwise compute the optimal solution through Algorithm 4.\\
  \hline
\end{tabular}\label{table:5}
\end{center}
\end{table}

}

\newpage

\newpage

\newpage

\newpage

\newpage

%

\newpage

\begin{thebibliography}{10}
\providecommand{\url}[1]{#1} \csname url@rmstyle\endcsname
\providecommand{\newblock}{\relax} \providecommand{\bibinfo}[2]{#2}
\providecommand\BIBentrySTDinterwordspacing{\spaceskip=0pt\relax}
\providecommand\BIBentryALTinterwordstretchfactor{4}
\providecommand\BIBentryALTinterwordspacing{\spaceskip=\fontdimen2\font
plus \BIBentryALTinterwordstretchfactor\fontdimen3\font minus
  \fontdimen4\font\relax}
\providecommand\BIBforeignlanguage[2]{{%
\expandafter\ifx\csname l@#1\endcsname\relax
\typeout{** WARNING: IEEEtran.bst: No hyphenation pattern has been}%
\typeout{** loaded for the language `#1'. Using the pattern for}%
\typeout{** the default language instead.}%
\else \language=\csname l@#1\endcsname \fi #2}}

\bibitem{FCC2003}
{Federal Communications Commission}, ``Facilitating opportunities
for flexible,
  efficient, and reliable spectrum use employing cognitive radio technologies,
  notice of proposed rule making and order, fcc 03-322,'' Dec. 2003.

\bibitem{Mitola1999}
J.~Mitola and G.~Q. Maguire, ``Cognitive radios: {Making} software
radios more
  personal,'' \emph{IEEE Personal Communications}, vol.~6, no.~4, pp. 13--18,
  Aug. 1999.

\bibitem{SimonKaykin:review}
S.~Haykin, ``Cognitive radio: {Brain-empowered} wireless
communications,''
  \emph{{IEEE} J. Select. Areas Commun.}, vol.~23, no.~2, pp. 201--202, Feb.
  2005.

\bibitem{Cui07:jsac}
Z.~Quan, S.~Cui, and A.~Sayed, ``Optimal linear cooperation for
spectrum
  sensing in cognitive radio networks,'' \emph{IEEE J. Select. Topics in Signal
  Processing}, vol.~2, no.~1, pp. 28--40, Feb. 2008.

\bibitem{Cui07:jsac1}
F.~Wang, M.~Krunz, and S.~Cui, ``Price-based spectrum management in
cognitive
  radio networks,'' \emph{IEEE J. Select. Topics in Signal Processing}, vol.~2,
  no.~1, pp. 74--87, Feb. 2008.

\bibitem{Gastpar_IEEE_TIT_2007}
M.~Gastpar, ``On capacity under receive and spatial spectrum-sharing
  constraints,'' \emph{{IEEE} Trans. Inform. Theory}, vol.~53, no.~2, pp.
  471--487, Feb. 2007.

\bibitem{Ghasemi_Sousa_IEEE_TWC_2007}
A.~Ghasemi and E.~S. Sousa, ``Fundamental limits of spectrum-sharing
in fading
  environments,'' \emph{{IEEE} Trans. Wireless Commun.}, vol.~6, no.~2, pp.
  649--658, Feb. 2007.

\bibitem{Liang:tradeoff08}
Y.-C. Liang, Y.~Zeng, E.~Peh, and A.~Hoang, ``Sensing-throughput
tradeoff for
  cognitive radio networks,'' \emph{{IEEE} Trans. Wireless Commun.}, vol.~7,
  no.~4, pp. 1326--1337, Apr. 2008.

\bibitem{GG02:eigenbeamforming}
S.~Zhou and G.~B. Giannakis, ``Optimal transmitter eigen-beamforming
and
  space-time block coding based on channel mean feedback,'' \emph{{IEEE} Trans.
  Signal Processing}, vol.~50, no.~10, pp. 2599--2613, Oct. 2002.

\bibitem{madhow:partialCSI2001}
E.~Visotsky and U.~Madhow, ``Space-time transmit precoding with
imperfect
  feedback,'' \emph{{IEEE} Trans. Inform. Theory}, vol.~47, no.~6, pp.
  2632--2639, Sept. 2001.

\bibitem{jafar:partialCSI2004}
S.~A. Jafar and A.~J. Goldsmith, ``Transmitter optimization and
optimality of
  beamforming for multiple antenna systems with imperfect feedback,''
  \emph{{IEEE} Trans. Wireless Commun.}, vol.~3, no.~4, pp. 1165--1175, July
  2004.

\bibitem{Aazhang:quantized03}
K.~K. Mukkavilli, A.~Sabharwal, E.~Erkip, and B.~Aazhang, ``On
beamforming with
  finite rate feedback in multiple antenna systems,'' \emph{{IEEE} Trans.
  Inform. Theory}, vol.~49, no.~10, pp. 2562--2579, Oct. 2003.

\bibitem{lan07:jsac}
L.~Zhang, Y.-C. Liang, and Y.~Xin, ``Joint beamforming and power
allocation for
  multiple access channels in cognitive radio networks,'' \emph{{IEEE} J.
  Select. Areas Commun.}, vol.~26, no.~1, pp. 38--51, Jan. 2008.

\bibitem{Liang:jstsp}
R.~Zhang and Y.-C. Liang, ``Exploiting multi-antennas for
opportunistic
  spectrum sharing in cognitive radio networks,'' \emph{IEEE J. Select. Topics
  in Signal Processing}, vol.~2, no.~1, pp. 88--102, Feb. 2008.

\bibitem{Liang:covariance01}
Y.-C. Liang and F.~P.~S. Chin, ``Downlink channel covariance matrix
({DCCM})
  estimation and its applications in wireless {DS-CDMA} systems,'' \emph{{IEEE}
  J. Select. Areas Commun.}, vol.~19, no.~2, pp. 222--232, Feb. 2001.

\bibitem{Jafar:uniview07}
S.~Srinivasa and S.~A. Jafar, ``The optimality of transmit
beamforming: a
  unified view,'' \emph{{IEEE} Trans. Inform. Theory}, vol.~53, no.~4, pp.
  1558--1564, Apr. 2007.

\bibitem{Boche:pcsi03}
E.~Jorswieck and H.~Boche, ``Optimal transmission with imperfect
channel state
  information at the transmit antenna array,'' \emph{Wireless Pers. Commun.},
  vol.~27, no.~1, pp. 33--56, Jan. 2003.

\bibitem{bental07:selected}
A.~Ben-Tal and A.~Nemirovski, ``Selected topics in robust convex
  optimization,'' \emph{Mathematical Programming}, vol.~1, no.~1, pp. 125--158,
  2007.

\bibitem{Boyd_optimization_book}
S.~Boyd and L.~Vandenberghe, \emph{Convex Optimization}.\hskip 1em
plus 0.5em
  minus 0.4em\relax Cambridge, UK: Cambridge University Press, 2004.

\bibitem{Rembert_opt_book}
R.~Reemtsen and J.-J. Ruckmann, \emph{Semi-Infinite
Programming}.\hskip 1em
  plus 0.5em minus 0.4em\relax Boston: Kluwer Academic Publishers, 1998.

\bibitem{Luo03:SPbstpaper}
S.~Vorobyov, A.~Gershman, and Z.-Q. Luo, ``Robust adaptive
beamforming using
  worst-case performance optimization: A solution to the signal mismatch
  problem,'' \emph{{IEEE} Trans. Signal Processing}, vol.~51, no.~2, pp.
  313--323, Feb. 2003.

\bibitem{Tomluo06:optimization}
Z.-Q. Luo and W.~Yu, ``An introduction to convex optimization for
  communications and signal processing,'' \emph{{IEEE} J. Select. Areas
  Commun.}, vol.~24, no.~8, pp. 1426--1438, Aug. 2006.

\bibitem{Sturm99:sedumi}
J.~F. Sturm, ``Using sedumi 1.02, a {MATLAB} toolbox for
optimization over
  symmetric cones,'' \emph{Optim. Meth. Softw.}, vol.~11, pp. 625--653, 1999.

\end{thebibliography}
\end{document}